\documentclass[aps,prb,twocolumn,showpacs]{revtex4}
\usepackage{graphicx}
\usepackage{color}
\usepackage{amsmath}
\newcommand{\tr}{{\bf 1}}
\usepackage{amsfonts}

\begin{document}

\title{Ashkin-Teller universality  in a  quantum double model of  Ising anyons}
\author{Charlotte Gils}
\affiliation{Theoretische Physik, ETH Zurich, 8093 Zurich, Switzerland}
\date{\today}

\begin{abstract}
We study a quantum double  model whose  degrees of freedom are  Ising anyons. 
The terms of the Hamiltonian of this system give rise to a competition 
 between single and double topologies.
 By studying  the energy spectra of the Hamiltonian at different values of the coupling constants,
  we find extended gapless regions which include a large number of critical points described by
conformal field theories with central charge $c=1$. 
These theories are part of the $\mathbb{Z}_2$ orbifold of the bosonic theory
 compactified on a circle.
We observe that the Hilbert space of our anyonic  model can be associated with extended Dynkin diagrams of affine Lie algebras 
which yields exact solutions at some critical points. In certain special regimes, our model corresponds to the Hamiltonian limit of
 the Ashkin-Teller model, and hence  integrability over a wide range of coupling parameters is established.

\end{abstract}


\maketitle


\section{Introduction}
 
There has been considerable  interest in 
emergent particles with fractional statistics, so-called anyons \cite{Leinaas_Myrheim_77,Froehlich_90}.
 Most prominently, anyons appear as quasiparticle excitations of the ground state  in the   
 fractional quantum Hall (FQH) liquids \cite{FQH_Laughlin,review_anyons}. 
 Anyons are also realized in quantum spin models in two spatial dimensions, 
  such as the  toric code model \cite{Kitaev_03}, the quantum dimer model on non-bipartite lattices
  \cite{Moessner_Sondhi_01}, and Kitaev's honeycomb model \cite{Kitaev_06}.
  The toric code model is a special case of a whole set of time-reversal and parity invariant 
   lattice models  that realize  doubled topological quantum field theories in (2+1) dimensions   
	 \cite{Levin_Wen_05,Fendley_Fradkin_05,Freedman_04}
    (`quantum double models').  
	The implementation of such models   in terms of lattice spin Hamiltonians
\cite{quantum_double1,quantum_double2,quantum_double3,optical_lattice}, or Josephson junction arrays \cite{Josephson_junction}
is under active investigation.
 
\begin{figure}[t]
\begin{center}
\includegraphics[width=8cm]{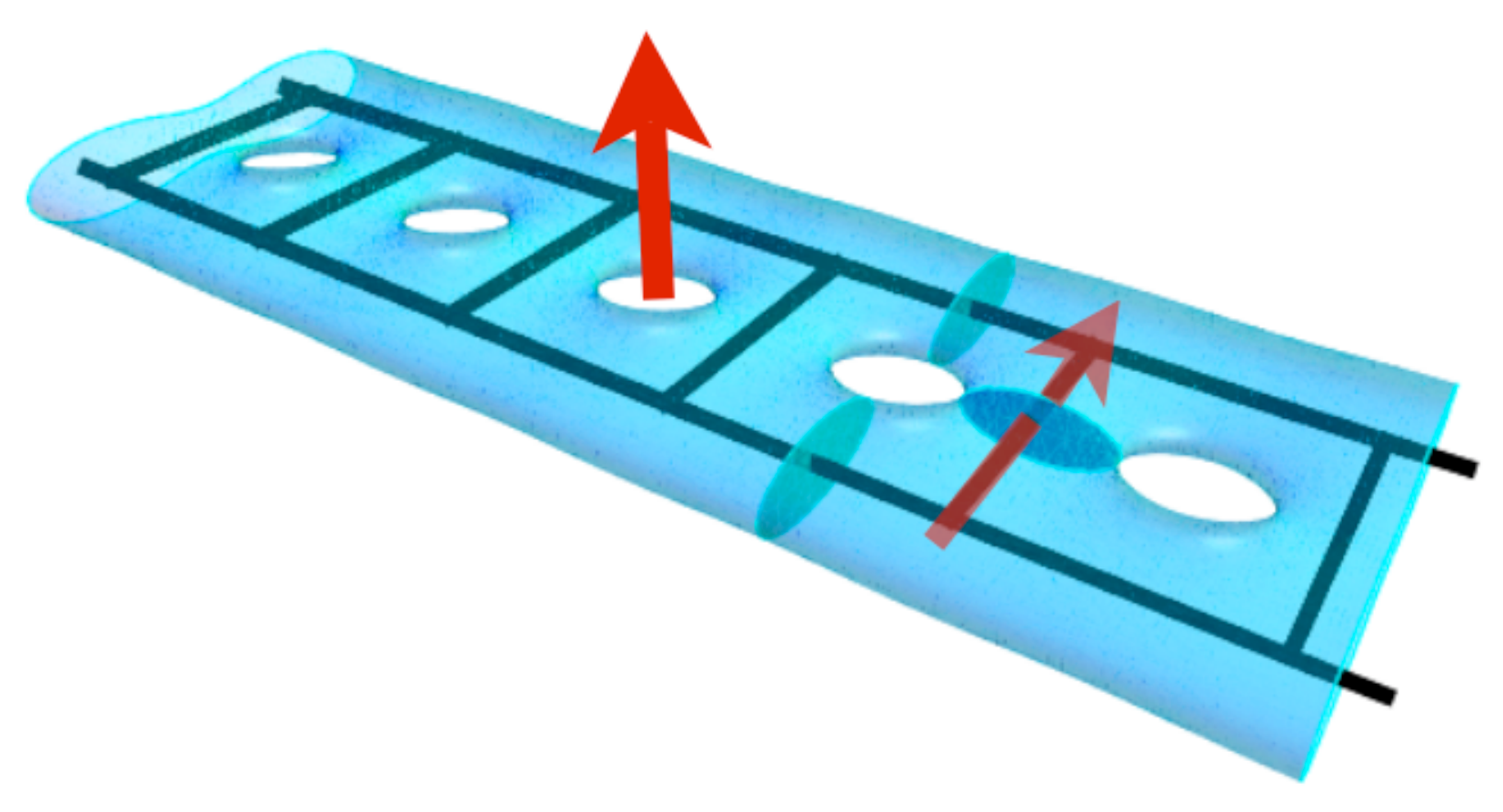}
\caption{The topology associated with our model of Ising anyons is a high-genus surface. The ladder skeleton inside the
sphere is a possible basis choice. In this basis, the two terms in the Hamiltonian
 project onto the `flux'  [$\tr$ (no flux), $\sigma$, or $\psi$]
  through the plaquettes (i.e., the holes of the
 high-genus surface) and the `flux' on the rungs of the ladder basis, respectively, as indicated by the red arrows.}
\label{high_genus1}
\end{center}
\end{figure}

  In this work, we are interested in the physics of a quantum double model	whose microscopic
   degrees of freedom 
   are non-abelian anyons.
   More specifically, we investigate a quantum double model whose degrees of freedom are Ising anyons.
    A simple example of a quantum double  model appeared in \cite{Levin_Wen_05,Fendley_Fradkin_05}  
 where the degrees of freedom are   Fibonacci anyons  \cite{Fibonacci} 
   located on the links of a honeycomb lattice. The   Hamiltonian 
  penalizes Fibonacci  anyon fluxes through the plaquettes of the lattice, 
   and it is exactly solvable.
This model has recently been considered on a
 ladder basis, where a competing term, which 
 penalizes Fibonacci anyons on the rungs of the 
  ladder basis, was added to the  Hamiltonian \cite{goldenladder}, see 
 Fig.~\ref{high_genus1}.
  In fact, the system studied in  \cite{goldenladder} is an example of a quantum double 
  model of non-abelian anyons with both a `string-net kinetic energy (plaquette fluxes) and a
   `string-net tension' (rung fluxes), as envisioned, but not studied,  in \cite{Levin_Wen_05}.
  It was found that the competition between the rung and plaquette fluxes 
		can be translated into the 
	 competition between two extreme topologies, each of them associated with a gapped phase. 
	 At equal magnitude of the coupling constants of rung and plaquette term, a critical point separating
	  the gapped phases was observed. This critical point, and a second critical phase, are described by certain conformal field 
	  theories. The Hilbert space of the model is associated with a $D_6$ Dynkin diagram which yields  
	   exact solutions at two critical points.
	   
	   In this paper, we focus on degrees of freedom corresponding to Ising anyons.
	     Ising anyons are currently the  most promising class of non-abelian anyons in the experimental context.
   A fractional quasiparticle charge of one quarter of the electron charge (as expected for the Ising anyon)
    has been measured  \cite{FQH_exp_Ising} which raises hopes that the quasiparticles in the
	 fractional quantum Hall state with filling fraction $\nu=5/2$ 
	  are indeed Ising anyons, as predicted in \cite{Moore_Read_91}.
Further systems with emergent Ising anyons are  $p+ip$ superconductors (or superfluids) \cite{Moore_Read_91,Read_Green_00}, and a 
	  quantum spin lattice model\cite{Kitaev_06}.
	  The Ising theory possesses an additional anyon species  as compared to the Fibonacci theory. There are  two types of particle species
	  (the Ising anyon $\sigma$, and the fermion $\psi$).
	  Hence
	  our model has two coupling parameters that can be tuned:
	   One of the coupling parameters varies the relative strength of rung and plaquette fluxes, 
	   while the other coupling parameter varies the relative strength of the 
	   Ising anyon and the  fermion (rung and plaquette) fluxes.
	 We  study the phase diagram as a function of the two coupling constants using exact diagonalization 
	    and analytical methods.
	We observe extended gapped and gapless phases, where the latter 
includes a number of 
	   critical theories which are described by  two-dimensional  rational conformal field theories (rCFT) with 
	    central charges $c=1$. 
	These critical points are part of the 	
		 $\mathbb{Z}_2$ orbifold of the bosonic theory compactified on  a circle of radius $R=\sqrt{2p}$ (where 
		 each integer $p>0$ gives rise to a separate rCFT). 
			The Hilbert space of our model is 
	   associated with the extended Dynkin diagrams $\hat{D}_4$ and $\hat{D}_6$ which yields exact solutions
	    a certain critical points. 
	Further results are established by identifying certain regimes of our model with the quantum Ashkin-Teller model.

The organization of this paper is as follows. After a  brief review of some essential properties of 
Ising anyons in section~\ref{Ising_anyons},
 we introduce our model in sections~\ref{hilbert_space} and \ref{hamiltonian}. We then present the results
  of an exact diagonalization study  of the
  Hamiltonian in section~\ref{num_results}. Thereafter, we present   exact solutions of our model at certain critical
   points, based on the association of the Hilbert space with certain extended Dynkin diagrams (section~\ref{exact_solutions}).

\section{Model}

\subsection{Ising anyons}\label{Ising_anyons}
In the following, we recapitulate some essential properties of the degrees of freedom of our model,
so-called Ising anyons \cite{Ising_anyons}.
There are three different particle `species' in the Ising theory, 
 the trivial particle $\tr$, the Ising anyon
 $\sigma$, and the  fermion $\psi$.
 
 The coupling of two Ising anyons is determined by the  fusion rules,
  which are the analogs of Clebsch-Gordon rules for ordinary angular momenta. The
   fusion rules of the Ising theory are given by
  \begin{equation}
\sigma\times \sigma= \tr+\psi\, ,\;\;\; \sigma\times \psi = \sigma\, ,\;\; \;\psi\times \psi = \tr\ ,\;\;\; \tr\times\tr =\tr\, .
\label{fusion_rules}
  \end{equation}
These fusion rules can be written in terms of the 
fusion matrices $N_j$ whose entries  $(N_j)^{j_1}_{j_2}$ 
 equal to one iff
the fusion of anyons of types $j_1$ and $j_2$ into $j$ is possible.
The fusion rules are related to the 
quantum dimensions $d_j$, $j=\tr,\sigma,\psi$,  by
   $N_j {\bf d}= d_j {\bf d}$, where  ${\bf d}$ is 
the eigenvector corresponding to the largest
positive 
eigenvalue of the matrix $N_j$.
The quantum dimensions of the Ising theory are 
 $d_{\tr}=1$, $d_{\sigma} = \sqrt{2}$ and $d_{\psi} = 1$, and 
 the total quantum dimension is $\mathcal{D}=\sqrt{d_{\tr}^2+d_{\sigma}^2+d_{\psi}^2} = 2$.

 In analogy to the $6j$-symbols for ordinary SU(2) spins, there exists a basis transformation $F$ 
that  relates the two differents ways three anyons $a$, $b$, and $c$ can fuse to a fourth anyon $d$,
\begin{equation}
\includegraphics[width=5.5cm]{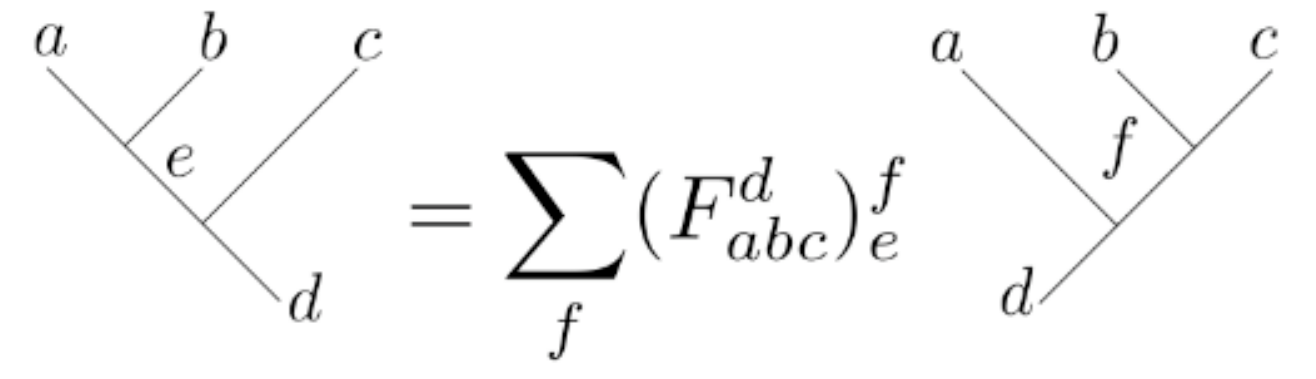}\, .
\label{Fmatrix_eq}
\end{equation}
Here, labels $a$, $b$,..., take values \tr, $\sigma$, and $\psi$,
 and the diagrams represent the quantum states of the `four anyon system' where anyons fuse in the
  specified order.
The non-trivial elements $(F_{abc}^d)_e^f$ (i.e., $(F_{abc}^d)_e^f\ne 1$) of the Ising theory are 
 $(F_{\sigma \psi \sigma}^{\psi})_{\sigma}^{\sigma}=-1$, and
\begin{equation}
F_{\sigma\sigma\sigma}^{\sigma}= \left ( \begin{array}{cc}(F_{\sigma\sigma\sigma}^{\sigma})_{\tr}^{\tr}&
(F_{\sigma\sigma\sigma}^{\sigma})_{\tr}^{\psi}\\(F_{\sigma\sigma\sigma}^{\sigma})_{\psi}^{\tr} &
(F_{\sigma\sigma\sigma}^{\sigma})_{\psi}^{\psi} 
\end{array}\right )= \frac{1}{\sqrt{2}} \left ( \begin{array}{cc}1 &1\\1&-1 \end{array} \right ) \, .
\label{Fmatrix_Ising}
\end{equation}

 The modular $S$-matrix  is a basis transformation which 
  relates the anyon `flux' of species $b$ through an anyon loop of species $a$
 to the case without anyon loop by
 \begin{equation}
\includegraphics[width=2.9cm]{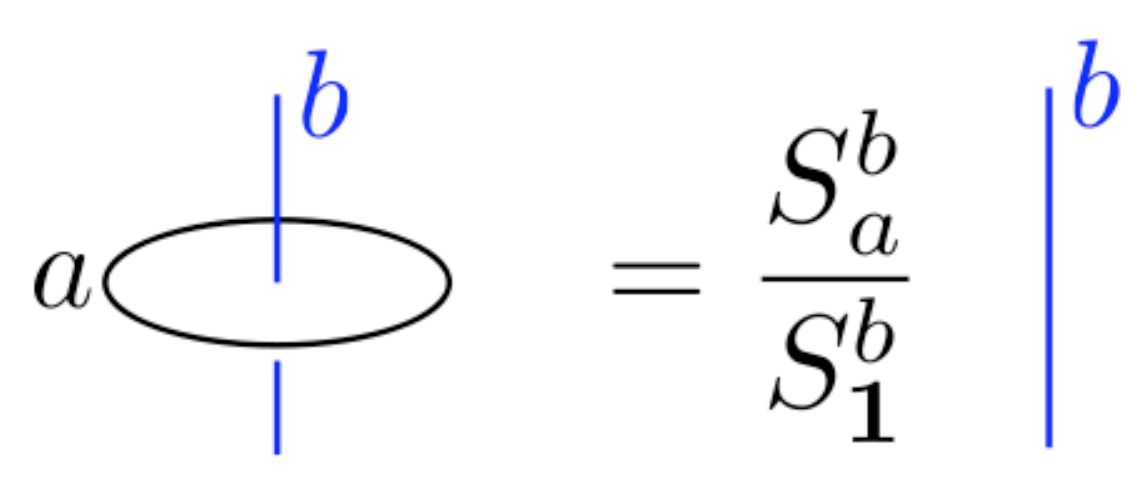}\, ,
 \label{S_eq}
 \end{equation}
and is of  form
\begin{equation}
S = \left ( \begin{array}{ccc}S_{\tr}^{\tr}&S_{\tr}^{\sigma}&S_{\tr}^{\psi}\\S_{\sigma}^{\tr}&S_{\sigma}^{\sigma}&S_{\sigma}^{\psi}\\
S_{\psi}^{\tr}&S_{\psi}^{\sigma}&S_{\psi}^{\psi} \end{array}\right )=\frac{1}{2}\left (\begin{array}{ccc}
 1&\sqrt{2}&1\\ \sqrt{2}&0&-\sqrt{2}\\1&-\sqrt{2}&1
\end{array}\right )\, ,
\label{Smatrix_Ising}
\end{equation}
for the case of Ising anyons.

 \begin{figure}[t]
\begin{center}
\includegraphics[width=5.5cm]{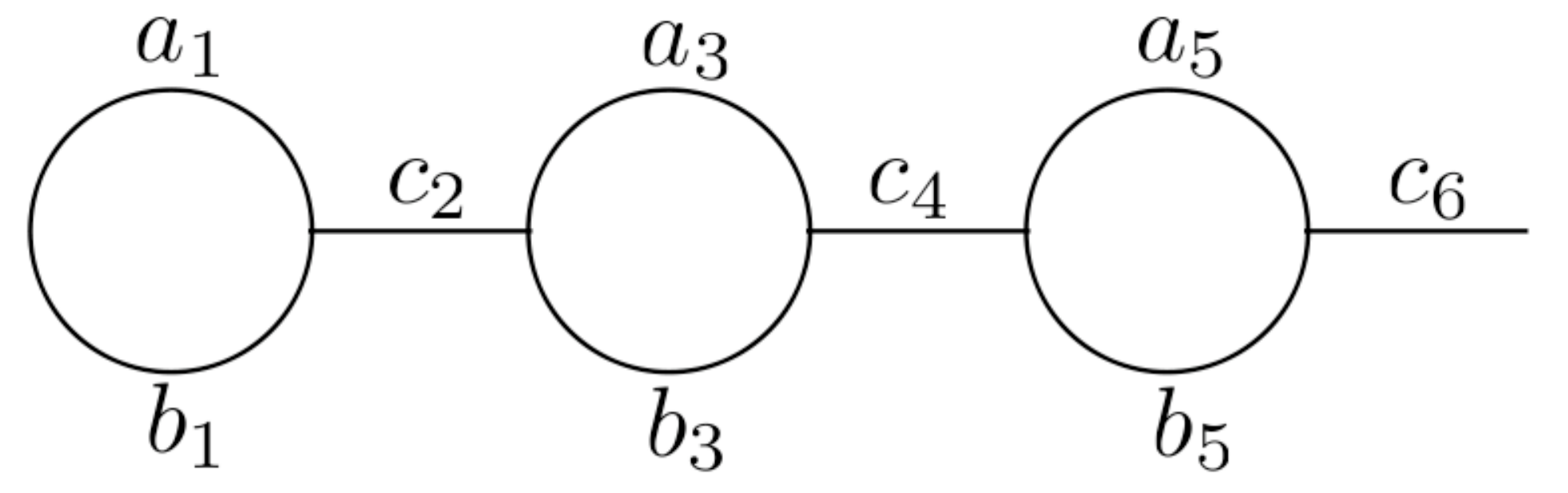}
\caption{Labeling of the basis (the fusion diagram). Periodic boundary conditions are applied, i.e.,
 $a_1=a_{2L+1}$, $b_1=b_{2L+1}$, and $c_2=c_{2L+2}$.
}
\label{basis_labeling}
\end{center}
\end{figure}  

\subsection{Hilbert space}\label{hilbert_space}
\begin{figure}[t]
\begin{center}
\includegraphics[width=8cm]{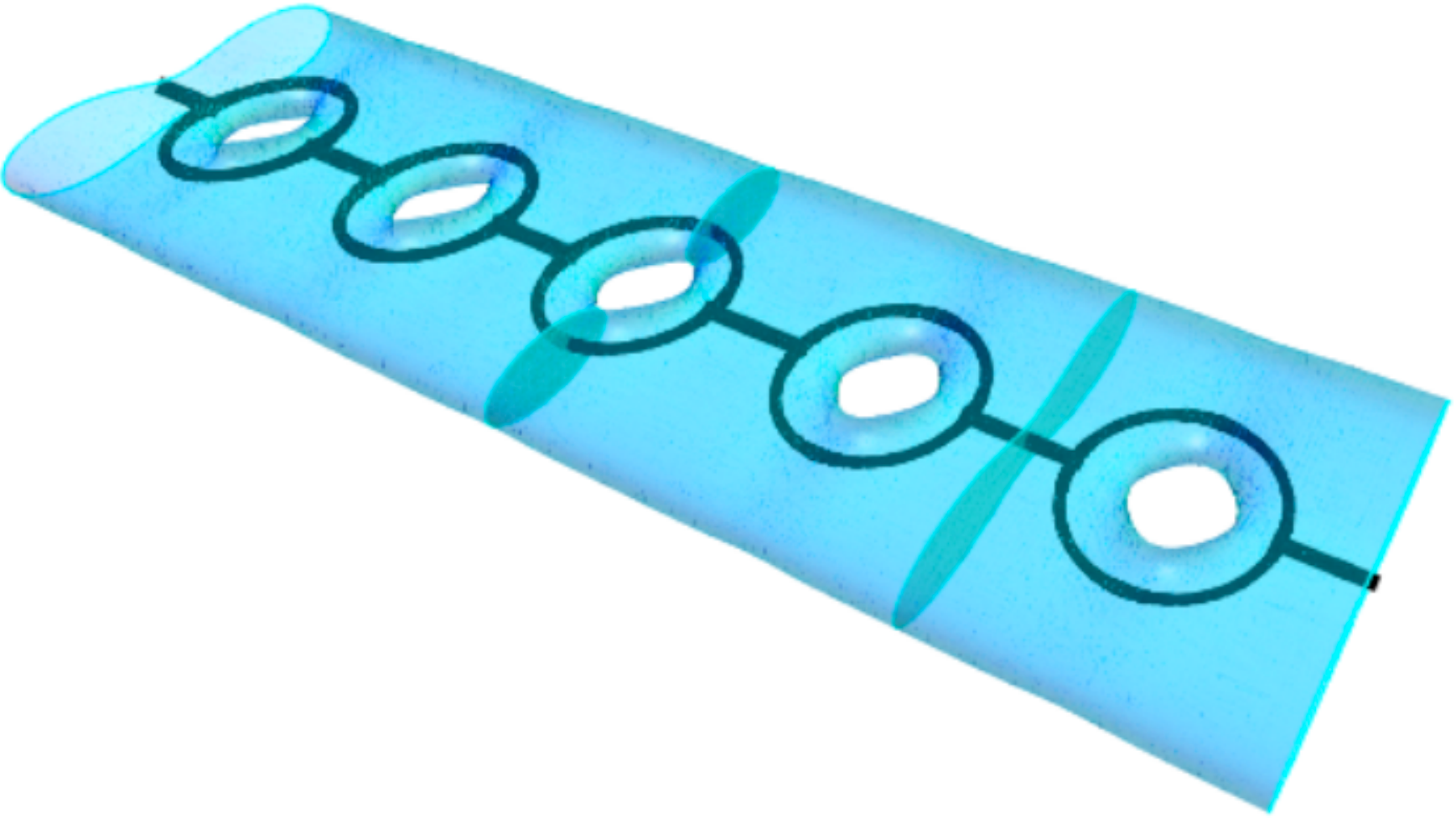}
\caption{A different decomposition of the high-genus
surface into three-puncture spheres (as compared to the one shown in 
Fig.~\ref{high_genus1}), 
 yields a different choice of basis (the black graph), also shown in Fig.~\ref{basis_labeling}.
}
\label{high_genus2}
\end{center}
\end{figure}

 Anyonic degrees of freedom are non-local, i.e., the Hilbert space of a 
  multi-anyon system is not the tensor product space of Hilbert spaces associated with 
  local  degrees of freedom, as is the case for ordinary spins.
  The Hilbert space of a multi-anyon system can be represented
  in terms of a fusion diagram which is a
   trivalent graph with each line segment symbolizing a certain anyon species, and the fusion
   rules being obeyed at the vertices (see Fig.~\ref{basis_labeling}).
  Each distinct occupation of the fusion diagram represents a basis state, and the inner product of
  two identical states is one, while the inner product of two  different states is zero.
By means of (for example) $F$-transformations, different basis choices of the same system can be  related.
  It is the topology (here, the high-genus surface of Figs.~\ref{high_genus1} and \ref{high_genus2}) that 
  defines the Hilbert space. 
Different basis choices correspond to different decompositions of the high-genus surface into three-punctured
 spheres, as can be seen by comparing Figs.~\ref{high_genus1} and \ref{high_genus2}.
We 
 formulate our Hamiltonian in the basis choice which is shown in Figs.~\ref{basis_labeling} and \ref{high_genus2}.
  
 In the terminology of Fig.~\ref{basis_labeling}, the 
 occupations of elements $c_i$ are either $c_i=\sigma$ ($i=2,4,...,2L$), 
 or $c_i\in\{\tr,\psi\}$ ($i=2,4,...,2L$).
 This means that there are two independent sectors of the Hilbert space of our model: 
\begin{itemize}
 \item IS (Integer sector): $c_i \in \{\tr,\psi\}$ ($i=2,4,...,2L$), $(a_i,b_i)\in \{(\tr,\tr),(\sigma,\sigma),(\psi,\psi),(\psi,\tr),(\tr,\psi)\}$
  ($i=1,3,...2L-1$).
 \item HIS (Half-integer sector): $c_i=\sigma$ ($i=2,4,...,2L$), $(a_i,b_i) \in \{(\tr,\sigma),(\sigma,\tr),(\psi,\sigma),(\sigma,\psi)\}$
  ($i=1,3,...,2L-1$).
\end{itemize}

Using the fusion matrices $N_j$, it is straightforward to evaluate the number of basis states, $B$, as a function of 
the number of plaquettes, $L$.
We apply periodic boundary conditions, i.e., $a_1=a_{2L+1}$, $b_1=b_{2L+1}$, $c_2=c_{2L+2}$.
The number of basis states is given by
\begin{eqnarray}
B &=& \sum_{\{a_i,b_i,c_i\}} (N_{c_2})_{a_1}^{b_1} (N_{c_2})^{a_3}_{b_3} (N_{c_4})^{b_3}_{a_3}
 ...(N_{c_{2L}})_{b_1}^{a_1} \\ 
  &=& \sum_{\{c_i\}} \prod_{i=1}^{L} {\rm Tr}(N_{c_{2i}}N_{c_{2i+2}}) \nonumber \\
  & =& \left \{ \begin{array}{ll} 
  \prod_{i=1}^L {\rm Tr}(N_{\sigma}N_{\sigma}) = 4^L & {\rm HIS}  \\[1mm]
  \sum_{\{c_i\in \{\tr,\psi\} \}} \prod_{i=1}^L {\rm Tr}(N_{c_{2i}}N_{c_{2i+2}}) = 4^L+2^L &{\rm IS,}\end{array}  \right. \nonumber 
\end{eqnarray}
where the summation $\sum_{\{a_i,b_i,c_i\}}$ runs over all possible labelings of the basis.

   \subsection{Hamiltonian}\label{hamiltonian}
The Hamiltonian contains two non-commuting terms which
  act in alternating manner  on even and odd labels $i$, $i=1,2,...,2L$ (terminology as
   in Fig.~\ref{basis_labeling}). The plaquette operator $P_i^{(s)}$
  projects onto  anyon flux $s$ (where $s\in \{\tr,\sigma,\psi\}$) through a plaquette indexed 
  by an odd integer $i$.
  The rung operator $R_i^{(s)}$ ($i$ even) projects onto the anyon occupation of a rung (i.e., it is diagonal in the 
   ladder basis
    of Fig.~\ref{high_genus1}, but not in the basis used [Fig.~\ref{basis_labeling}]).

   In the most general form, the Hamiltonian is given by
   \begin{eqnarray*}
H &= & -J_p \sum_{i=1}^L\left ( J_{\tr} P^{(\tr)}_{2i-1} +J_{\sigma}  P_{2i-1}^{(\sigma)} 
+J_{\psi}  P_{2i-1}^{(\psi)} \right ) 
\nonumber \\ 
&&- J_r \sum_{i=1}^L\left ( J_{\tr} R^{(\tr)}_{2i} +J_{\sigma} R_i^{(\sigma)}
+J_{\psi}  R_{2i}^{(\psi)}\right ) \, .
\label{Hamiltonian}
\end{eqnarray*}
We set 
$J_p=\cos(\theta)$, $J_r=\sin(\theta)$, $J_{\tr}=\cos(\phi)$, $J_{\psi}=\sin(\phi)$ and rewrite the Hamiltonian as
 (note that $P_i^{(\tr)}+P_i^{(\sigma)}+P_i^{(\psi)}=1$, and $R_i^{(\tr)}+R_i^{(\sigma)}+R_i^{(\psi)}=1$),
	\begin{eqnarray}
	 H&=& -\cos(\theta) \sum_{i=1}^L P_{2i-1}-\sin(\theta)\sum_{i=1}^{L} R_{2i} \label{Ham}\\ 
	P_i&=&  \cos(\phi) P^{(\tr)}_{i}+\sin(\phi) P_{i}^{(\psi)} \nonumber \\
    R_i&=& \cos(\phi)R_{i}^{(\tr)}+\sin(\phi) R_{i}^{(\psi)} \nonumber 
	 \end{eqnarray}  
The parameter $\theta$ controls the dimerization of the model. If $J_r=J_p$, i.e., $\theta=\pi/4$ or $\theta=5\pi/4$,
 the  dimerization is zero, i.e., the local terms $H_i$ (where
  $H_i=P_i$ if $i$ odd, and $H_i=R_i$ if $i$ even)  have
  identical coupling strengths
  at each `site' $i$.
 
  \begin{figure}
  \begin{center}
  \includegraphics[width=7.7cm]{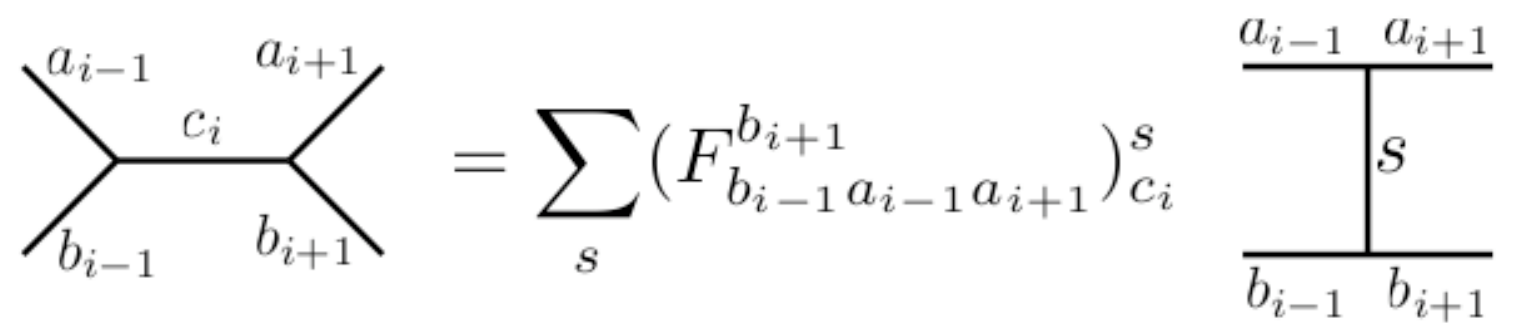}
  \caption{$F$-transformation of a local element of the basis of Fig.~\ref{basis_labeling} to the ladder basis.
  }\label{rung_term}
  \end{center}
  \end{figure}

  The exact form of the terms $P_i^{(s)}$ and $R_i^{(s)}$ was  
   discussed in  \cite{goldenladder} for the case of Fibonacci anyon degrees of freedom. However, 
   we  shall repeat the derivation of this non-standard Hamiltonian for the case of Ising anyons.
 We begin with the local plaquette term $P_i^{(s)}$.
We insert  an additional anyon loop of type $t\in\{1,\sigma,\psi\}$ into the center of the plaquette composed by
 variables $(a_i,b_i)$, 
 and project onto
the flux through this additional loop (and hence the flux through the plaquette)
  using the $S$-matrix Eq.~(\ref{S_eq}),
\begin{equation}
P_i^{(s)}\left | \parbox{1.7cm}{\includegraphics[width=1.8cm]{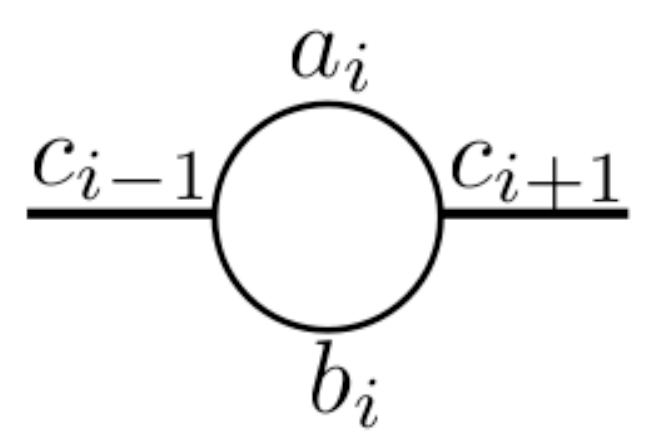}} \right \rangle
 = \sum_{t=\tr,\sigma,\psi} S_{\tr}^s S_t^s \left | \parbox{1.7cm}{\includegraphics[width=1.8cm]{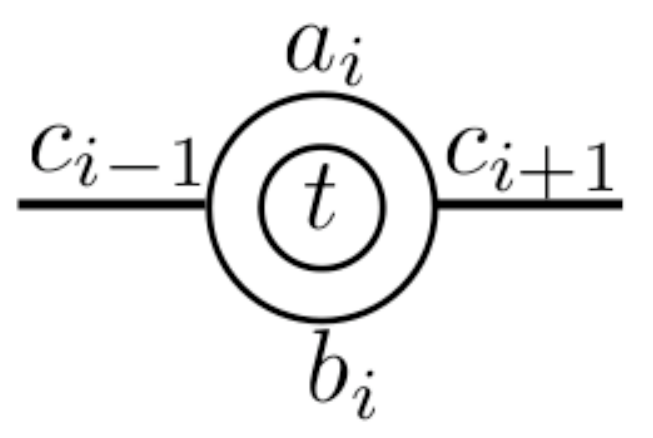}} \right \rangle \, .
 \end{equation}
 We proceed further as follows,
 \begin{eqnarray}
 && \left | \parbox{1.7cm}{\includegraphics[width=1.8cm]{plaqterm2.pdf}} \right \rangle=
 \sum_{a_i'} (F_{a_ia_it}^t)_{\tr}^{a_i'} 
 \left | \parbox{1.7cm}{\includegraphics[width=1.8cm]{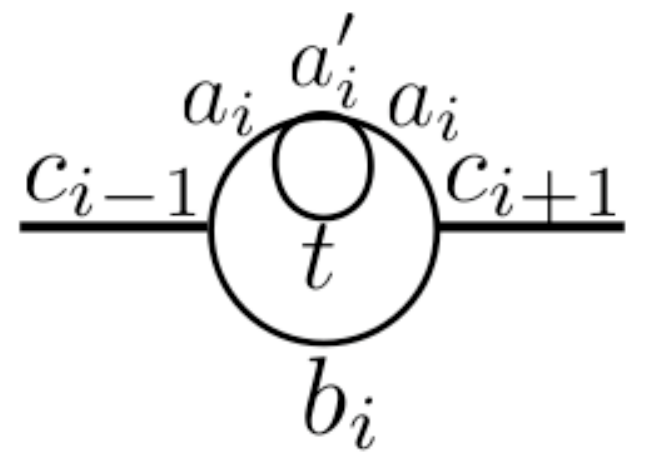}} \right \rangle \nonumber \\
 & =&  \sum_{a_i',b_i'} (F_{a_ia_it}^t)_{\tr}^{a_i'} (F_{c_{i+1}b_it}^{a_i'})_{a_i}^{b_i'}
 \left | \parbox{1.7cm}{\includegraphics[width=1.8cm]{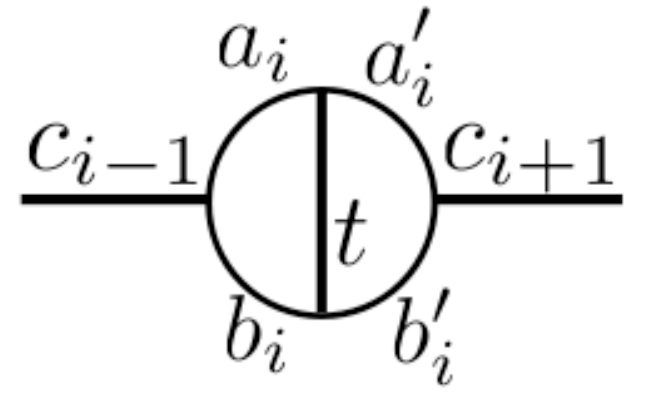}} \right \rangle \nonumber \\
 & =&  \sum_{a_i',b_i',m} (F_{a_ia_it}^t)_{\tr}^{a_i'} (F_{c_{i+1}b_it}^{a_i'})_{a_i}^{b_i'}
 (F_{c_{i-1}a_it}^{b_i'})_{b_i}^m  
 \left | \parbox{1.7cm}{\includegraphics[width=1.8cm]{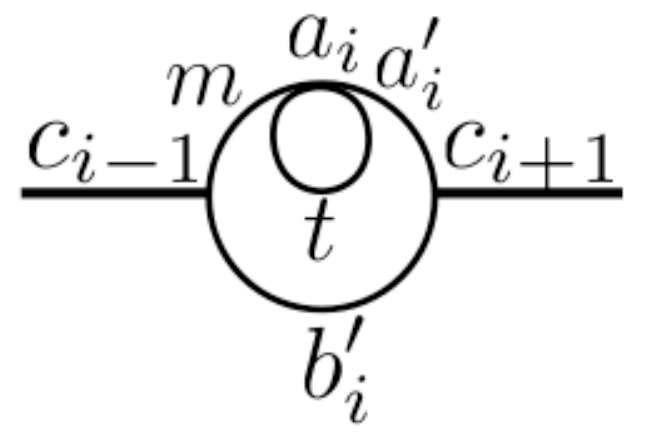}} \right \rangle  \nonumber \\
 &=& \sum_{a_i',b_i'}  (F_{c_{i+1}b_it}^{a_i'})_{a_i}^{b_i'}
 (F_{c_{i-1}a_it}^{b_i'})_{b_i}^{a_i'}
 \left | \parbox{1.7cm}{\includegraphics[width=1.8cm]{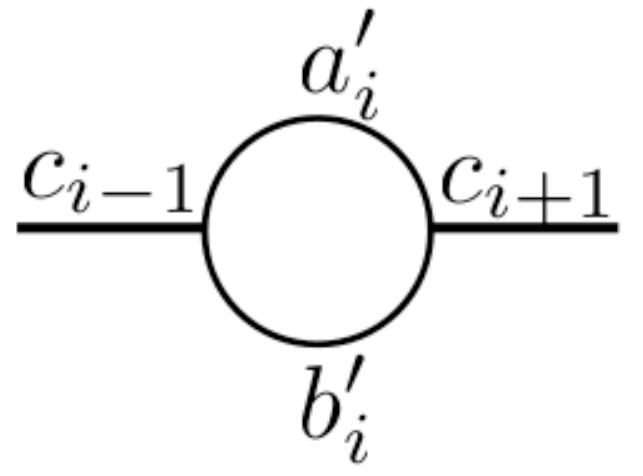}} \right \rangle\, ,  
 \label{local_plaq_term}
\end{eqnarray}
where we used the identity
\begin{equation}
\includegraphics[width=3.8cm]{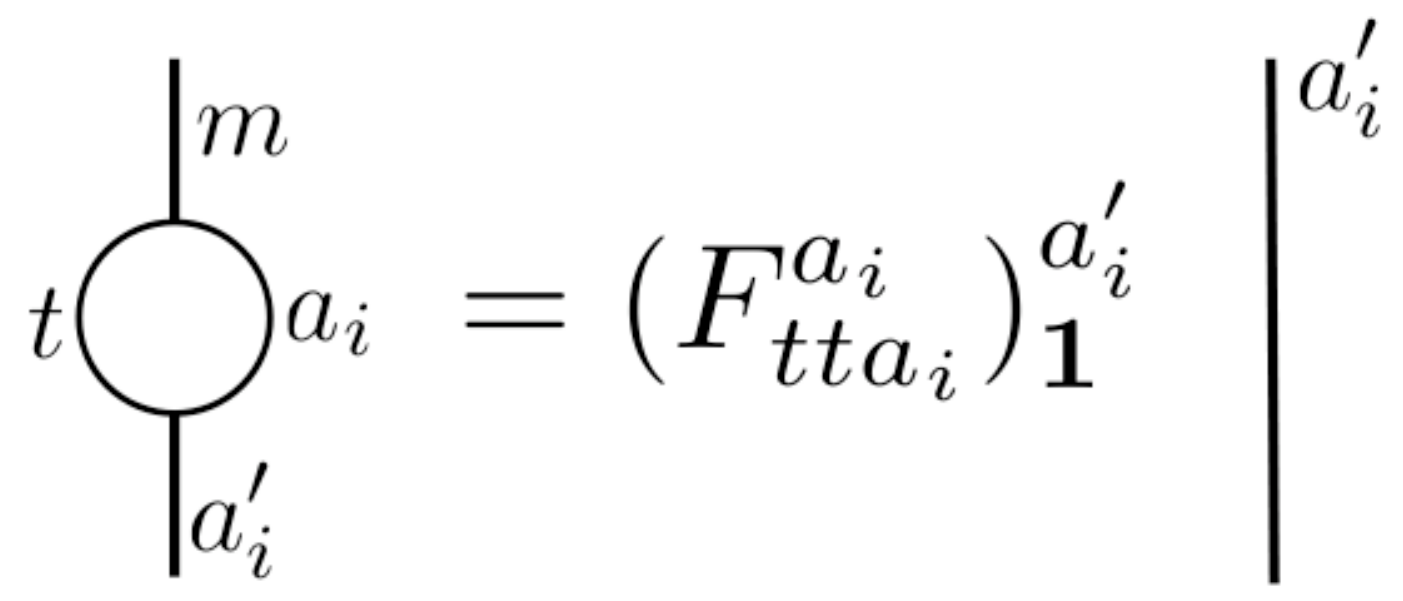}\, ,
\end{equation}
 and the orthogonality relation $\sum_e (F_{abc}^d)_f^e (F_{dab}^c)_e^k=\delta_{e,k}$.

By using an $F$-transformation, it is possible to transform between the  basis of Fig.~\ref{basis_labeling} 
 and the  ladder basis, as shown for a local element in Fig.~\ref{rung_term}.
 Using such a transformation, the projector onto a rung with occupation $s$ is given by
\begin{equation}
R_i^{(s)}|c_i\rangle = \sum_{c'_i} (F_{b_{i-1}a_{i-1}a_{i+1}}^{b_{i+1}})_{c_i}^s
(F_{b_{i-1}a_{i-1}a_{i+1}}^{b_{i+1}})_{c'_i}^s  |c'_i\rangle \,.
\label{local_rung_term}
 \end{equation}
 
 It is straightforward to construct a matrix representation of the Hamiltonian (\ref{Ham}). 
 In the half-integer sector (HIS), the variables at even sites $i$ are fixed, i.e., $c_i=\sigma$. We 
   associate the local `site' variables $(a_i,b_i) \in \{ (\tr,\sigma)$, $(\sigma,\tr)$, $(\psi,\sigma)$, $(\sigma,\psi)\} $  ($i$ odd)
   with the four
  unit vectors in four dimensions, respectively, and define $n^{(\tr,\sigma)}={\rm Diag}(1,0,0,0)$, $n^{(\sigma,\tr)}={\rm Diag}(0,1,0,0)$,
  $n^{(\psi,\sigma)}={\rm Diag}(0,0,1,0)$, and $n^{(\sigma,\psi)}={\rm Diag}(0,0,0,1)$.
Evaluating Eqs.~(\ref{local_plaq_term}) and (\ref{local_rung_term}) using  
  the $F$- and $S$-matrix elements (see section~\ref{Ising_anyons}) yields a $4\times 4$ representation of the
  Hamiltonian in the HI sector,
  \begin{widetext}
 \begin{eqnarray}
  H^{\rm HIS} &=& -\cos(\theta)\sum_{i\ {\rm odd}} \{\cos(\phi) B_i^{\tr}+\sin(\phi)B_i^{\psi}\} 
  -\sin(\theta)  \cos(\phi) \sum_{i\ {\rm even}} \{n^{(\tr,\sigma)}_{i-1}n^{(\tr,\sigma)}_{i+1}+n^{(\sigma,\tr)}_{i-1}n^{(\sigma,\tr)}_{i+1}
  + n^{(\psi,\sigma)}_{i-1} n^{(\psi,\sigma)}_{i+1}+n^{(\sigma,\psi)}_{i-1}n^{(\sigma,\psi)}_{i+1}\} \nonumber \\
	 &&-\sin(\theta)\sin(\phi) \sum_{i \ {\rm even}} \{ n^{(\tr,\sigma)}_{i-1}n^{(\psi,\sigma)}_{i+1}+ 
	n^{(\sigma,\tr)}_{i-1}n^{(\sigma,\psi)}_{i+1}+n^{(\psi,\sigma)}_{i-1}n^{(\tr,\sigma)}_{i+1}+n^{(\sigma,\psi)}_{i-1}n^{(\sigma,\tr)}_{i+1}\},
\label{Ham_HIS}
  \end{eqnarray}
  where
  \begin{eqnarray*}
B^{\tr} &=& \frac{1}{4}\left ( \begin{array}{cccc}1&1&1&1\\1&1&1&1\\1&1&1&1\\1&1&1&1\end{array}\right ),\hspace{1cm}
B^{\psi} = \frac{1}{4}\left ( \begin{array}{rrrr}1&-1&1&-1\\-1&1&-1&1\\1&-1&1&-1\\-1&1&-1&1\end{array}\right ).
\end{eqnarray*}
The Hamiltonian (\ref{Ham_HIS}) is invariant under variable exchanges $(\tr,\sigma)\leftrightarrow (\psi,\sigma)$ and 
$(\sigma,\tr)\leftrightarrow (\sigma,\psi)$ (independently and simultaneously). It is also invariant under simultaneous exchange of 
 variables  $(\tr,\sigma)\leftrightarrow (\sigma,\tr)$
 and $(\psi,\sigma)\leftrightarrow (\sigma,\psi)$.

In a similar manner as for the half-integer sector, it is possible to construct a $7 \times 7$ 
 matrix representation of the Hamiltonian (\ref{Ham}) in the integer sector. Let
 the variables $\tr,(\tr,\tr),(\tr,\psi),(\sigma,\sigma),(\psi,\tr),(\psi,\psi),\psi$ be associated with 
 the seven unit vectors in seven dimensions (in this order) and let $m^{\tr}={\rm Diag}(1,0,0,0,0,0,0)$, $m^{(\tr,\tr)}={\rm Diag}(0,1,0,0,0,0,0)$, and so on.
 In this notation, the Hamiltonian (\ref{Ham}) in the integer sector takes the form
 \begin{eqnarray}
H^{\rm IS} &=& -\cos(\theta)\sum_{i \ {\rm odd} } \{\cos(\phi)[ m^{\tr}_{i-1} M^{\tr,\tr}_i m^{\tr}_{i+1} + m_{i-1}^{\psi}
 M^{\psi,\tr}_i m_{i+1}^{\psi}]
 +\sin(\phi) [m_{i-1}^{\tr} M_i^{\tr,\psi} m_{i+1}^{\tr} +m_{i-1}^{\psi} M_i^{\psi,\psi } m_{i+1}^{\psi}]\} \nonumber \\
&& -\sin(\theta) \cos(\phi) \sum_{i \ {\rm even}}  \{ m_{i-1}^{(\tr,\tr)} m_{i+1}^{(\tr,\tr)}
+ m_{i-1}^{(\psi,\psi)} m_{i+1}^{(\psi,\psi)} + m_{i-1}^{(\sigma,\sigma)} M^{\sigma}_i  m_{i+1}^{(\sigma,\sigma)}\} \nonumber \\
&&-\sin(\theta)\sin(\phi) \sum_{i \ {\rm even}} \{  m_{i-1}^{(\tr,\psi)} m_{i+1}^{(\psi,\tr)}+ m_{i-1}^{(\psi,\tr)} m_{i+1}^{(\tr,\psi)}
+ m_{i-1}^{(\tr,\tr)} m_{i+1}^{(\psi,\psi)}+ m_{i-1}^{(\psi,\psi)} m_{i+1}^{(\tr,\tr)} \} ,
   \label{Ham_IS}
\end{eqnarray}
where
\begin{eqnarray*}
M^{\tr,\tr} &=& \frac{1}{4} \left ( \begin{array}{ccccccc} 0&0&0&0&0&0&0\\0&1&0&\sqrt{2}&0&1&0\\0&0&0&0&0&0&0
\\ 0 &\sqrt{2}&0&2 &0&\sqrt{2}&0\\0&0&0&0&0&0&0\\0&1&0&\sqrt{2}&0&1&0\\0&0&0&0&0&0&0 \end{array}\right ),  \hspace{0.5cm}
M^{\psi,\tr} =  \frac{1}{4} \left ( \begin{array}{ccccccc} 0&0&0&0&0&0&0\\0&0&0&0&0&0&0\\
0&0&1&\sqrt{2}&1&0&0\\ 0 &0&\sqrt{2}&2 &\sqrt{2}&0&0\\0&0&1&\sqrt{2}&1&0&0\\0&0&0&0&0&0&0\\0&0&0&0&0&0&0 \end{array}\right ), 
\hspace{0.5cm}
M^{\sigma,\sigma} = \frac{1}{2} \left ( \begin{array}{ccccccc} 1&0&0&0&0&0&1\\0&0&0&0&0&0&0\\0&0&0&0&0&0&0\\
0&0&0&0&0&0&0\\0&0&0&0&0&0&0\\0&0&0&0&0&0&0\\1&0&0&0&0&0&1\end{array}\right ),
\end{eqnarray*}
\begin{eqnarray*}
M^{\tr,\psi} &=& \frac{1}{4} \left ( \begin{array}{ccccccc} 0&0&0&0&0&0&0\\0&1&0&-\sqrt{2}&0&1&0\\0&0&0&0&0&0&0
\\ 0 &-\sqrt{2}&0&2 &0&-\sqrt{2}&0\\0&0&0&0&0&0&0\\0&1&0&-\sqrt{2}&0&1&0\\0&0&0&0&0&0&0 \end{array}\right ),   \hspace{1cm}
M^{\psi,\psi} =  \frac{1}{4} \left ( \begin{array}{ccccccc} 0&0&0&0&0&0&0\\0&0&0&0&0&0&0\\
0&0&1&-\sqrt{2}&1&0&0\\ 0 &0&-\sqrt{2}&2 &-\sqrt{2}&0&0\\0&0&1&-\sqrt{2}&1&0&0\\0&0&0&0&0&0&0\\0&0&0&0&0&0&0 \end{array}\right ) .
\end{eqnarray*}
\end{widetext}
Hamiltonian (\ref{Ham_IS}) is invariant under variables exchanges (i) $(\tr,\tr)\leftrightarrow (\psi,\psi)$
 and (ii) $(\tr,\psi)\leftrightarrow (\psi,\tr)$ (independently and simultaneously). It is also invariant under  simultaneous exchange of 
 $\tr\leftrightarrow \psi$, $(\psi,\tr)\leftrightarrow (\psi,\psi)$
  and $(\tr,\psi)\leftrightarrow (\tr,\tr)$ (where exchanges (i) and (ii) are applicable, too).


\subsection{The half-integer sector and the  quantum Ashkin-Teller model}\label{AT_HIS}
 The half-integer sector of the high-genus ladder of Ising anyons studied in this paper 
 is equivalent to the quantum Ashkin-Teller 
 model. The quantum Ashkin-Teller model, which can also be mapped onto the staggered XXZ chain,
  is the Hamiltonian limit of the classical 
 Ashkin-Teller model,  
 and it was first studied in \cite{Kohmoto}.

  The correspondence of our model in the half-integer sector and the quantum Ashkin-Teller model
   becomes immediately apparent when relabeling the indices of Hamiltonian Eq.~(\ref{Ham_HIS}) according to $i-1\to n$, $i+1\to n+1$, and 
    comparing this Hamiltonian 
   with the Hamiltonian given in \cite{Kohmoto} (see also \cite{Gehlen}).
  Relating the coupling constants $\theta$ and $\phi$ to the ones in \cite{Kohmoto} 
  allows us to confirm the results of sections~\ref{Num_HIS} and \ref{Exact_HIS}, and 
   add further details to the phase diagram Fig.~\ref{PD_theta25_phi_HIS}.
We believe that the numerical results discussed section~\ref{Num_HIS} are of interest despite prior studies of
the quantum Ashkin-Teller model, 
 and we note that the derivation of the exact solution  in section~\ref{Exact_HIS} is 
  a consequence of the  unique structure of the Hilbert space of our anyonic model.

\subsection{Numerical method }
We diagonalize the Hamiltonian matrix
 using the Lanczos algorithm \cite{Lanczos}.
  By employing periodic boundary conditions, we obtain the energy eigenvalues
   as a function of momenta  $k_x=2\pi n/L$, $n=1,2,...,L$, as well 
   as $k_y=0,\pi$ (invariance of the Hamiltonian under exchange of the $a_i$ and $b_i$ variables, this symmetry
    corresponds to simultaneous variable exchanges  $(\tr,\sigma)\leftrightarrow (\sigma,\tr)$ and $(\psi,\sigma)\leftrightarrow (\sigma,\psi)$ in
	 the half-integer sector, and exchange $(\tr,\psi)\leftrightarrow (\psi,\tr)$ in the integer sector).
     We employ an implementation of the Lanczos algorithm in the ALPS library \cite{alps}.



\section{Numerical Results}\label{num_results}
We first outline the topological feature of our model that 
determines its criticality
 at equal magnitude of rung  and plaquette term.
 We recapitulate the identification of a conformal field theory
 based on the energy spectrum in a system of finite size  
and  review the   operator content of the $\mathbb{Z}_2$ orbifold of the compactified bosonic theory.
Then, we present the results  of the exact diagonalization of
 the Hamiltonian matrix.

\subsection{Competing topologies}

 The competition between the rung and plaquette terms correspond to a competition between single and double topologies
  \cite{goldenladder}.
 This can be understood by switching to the ladder basis of Fig.~\ref{high_genus1}.
  In the ladder basis, the  plaquette term projects onto the flux $s$ through the plaquette, and the  rung term 
		 projects onto the flux $s$ on the rung of the ladder.
 We consider  the Hamiltonian at the points $J_p=1$, $J_r=0$, $J_{\psi}=0$ ($\theta=0$, $\phi=0$), and
    $J_r=1$, $J_p=0$, $J_{\psi}=0$ ($\theta=\pi/2$, $\phi=0$),
    respectively.
For the former choice of coupling constants,
 the rung term is zero, and the Hamiltonian  favors the absence of $\sigma$- and $\psi$-fluxes through the
 plaquettes.
 However, if there are no fluxes through the holes of the high-genus surface, the holes can be closed, and we are left with 
  a single cylinder (a torus for the case of peridic boundary conditions). 
In contrast, at the latter choice of coupling parameters, 
 the plaquette term is zero, the Hamiltonian favors the absence of $\sigma$ and $\psi$ particles on
  the rungs. Hence,  the rungs can be `cut off', and the resulting surface is that of two independent cylinders (two tori for 
  periodic boundaries).
  In this work, we mainly consider the points of equal magnitude of rung and plaquette term where the competition
    between single  and double topologies renders the system critical over a large range of coupling parameters.


\begin{table}[t]
\begin{tabular}{c||c|c|c|c|c|c|c|c} 
\hspace{1cm}$p$&$36$ &$16$ &$9$  &$6$ & $4$  & $3$&  $2$ & $1$ \\
& && & scft& Potts & para &Ising$^2$ & KT\\
$h_0+\bar{h}_0$& && & & & & & 
 \\ \hline \hline\hline
$\frac{1}{8}$&$\frac{1}{8}$&  $\frac{1}{8}$&$\frac{1}{8}$ & $\frac{1}{8}$& $\frac{1}{8}$& $\frac{1}{8}$& $\frac{1}{8}$ & $\frac{1}{8}$  \\ \hline \hline
$\frac{1}{8}$&$\frac{1}{8}$&$\frac{1}{8}$ &$\frac{1}{8}$ &$\frac{1}{8}$& $\frac{1}{8}$&$\frac{1}{8}$&$\frac{1}{8}$&$\frac{1}{8}$  \\ \hline \hline
$\frac{9}{8}$&$\frac{9}{8}$&$\frac{9}{8}$ &$\frac{9}{8}$&$\frac{9}{8}$&$\frac{9}{8}$&$\frac{9}{8}$&$\frac{9}{8}$&$\frac{9}{8}$\\ \hline \hline
$\frac{9}{8}$&$\frac{9}{8}$&$\frac{9}{8}$ &$\frac{9}{8}$ &$\frac{9}{8}$&$\frac{9}{8}$&$\frac{9}{8}$&$\frac{9}{8}$&$\frac{9}{8}$ \\ \hline \hline  
$2$& $2$ & $2$ &$2$ & $2$  &$2$ & $2$ &  $2$ & $2$  \\ \hline \hline 
$\frac{1}{2p}$&$\frac{1}{72}$& $\frac{1}{32}$ & $\frac{1}{18}$& $\frac{1}{12}$ & $\frac{1}{8}$ &$\frac{1}{6}$&$\frac{1}{4}$&- \\ \hline \hline
$\frac{4}{2p}$&$\frac{1}{18}$&$\frac{1}{8}$& $\frac{2}{9}$ &$\frac{1}{3}$&$\frac{1}{2}$&$\frac{2}{3}$&-&-    \\ \hline  \hline 
 $\frac{9}{2p}$&$\frac{1}{8}$&$\frac{9}{32}$&   $\frac{1}{2}$&$\frac{3}{4}$ &$\frac{9}{8}$ &-&-&-  \\ \hline \hline 
 $\frac{16}{2p}$&$\frac{2}{9}$& $\frac{1}{2}$& $\frac{8}{9}$ & $\frac{4}{3}$&-&-& -& -    \\ \hline  \hline 
 $\frac{25}{2p}$&$\frac{25}{72}$&$\frac{25}{32}$& $\frac{25}{18}$ &$\frac{25}{12}$&-&-&-  & -  \\ \hline \hline 
 $\frac{36}{2p}$&$\frac{1}{36}$&$\frac{9}{8}$&$2$ &- &- &- &-  \\ \hline \hline 
 $\frac{49}{2p}$&$\frac{49}{72}$& $\frac{49}{32}$&$\frac{49}{18}$&-&-&-&-&- \\ \hline \hline
 $\frac{64}{2p}$&$\frac{8}{9}$& $2$&$\frac{32}{9}$&-&-&-&-&-\\ \hline \hline
 $\frac{p}{2}$&$18 $ &$8$& $\frac{9}{2}$   & $3$   &  $2$ & $\frac{3}{2}$ & $1$    & $\frac{1}{2}$  \\ \hline  \hline
 $\frac{p}{2}$&$18$ &$8$& $\frac{9}{2}$   & $3$   &  $2$ & $\frac{3}{2}$ & $1$    & $\frac{1}{2}$   \\ 
\end{tabular}
\caption{Scaling dimensions $h_0+\bar{h}_0$ (aside from the ground state with $h=\bar{h}=0$) of the 
operators of the $\mathbb{Z}_2$ orbifold of the boson compactified on a circle
 of radius $ R=\sqrt{2p}$ for some 
theories (i.e., some integer $p>0$).
  The scaling dimensions which depend on $p$ are given by  $h+\bar{h}_0=\frac{n^2}{2p}$ where $n=1,...,p-1$, and by $p/2$.
 The following  abbreviations are used: scft = superconformal CFT with $c=1$, Potts = $4$-state Potts theory, para = parafermion CFT
  with $c=1$,
  Ising$^2$ = square of the Ising CFT, KT = Kosterlitz-Thouless transition.
 }
 \label{comp_boson_table}
\end{table}

\subsection{Identification of conformal field theories}
The spectrum of a conformal field theory (CFT) in a system of finite size $L$ and periodic boundary
 conditions has the following energy eigenvalues\cite{Cardy}
\begin{equation}
E = E_1L +\frac{2\pi v}{L} \left (-\frac{c}{12}+ h+\bar{h} \right ),
\label{CFT_energy_levels}
\end{equation}
where $c$ is the central charge of the CFT, and the velocity $v$ is an 
overall scale factor.
The scaling dimensions $h+\bar{h}$ take the form $h=h^0+n$, $\bar{h}=\bar{h}^0+\bar{n}$,
where $n$ and $\bar{n}$ are non-negative integers, and $h^0$ and $\bar{h}^0$ are the holomorphic and antiholomorphic 
conformal weights of primary fields of a given CFT of central charge $c$. Energies with $h$ and $\bar{h}$ such that
 $n$ and $\bar{n}$ zero  are associated with primary fields while energies with $n$ and/or $\bar{n}$ non-zero 
  correspond to descendant fields.
There are some constraints on the  momenta $k_x$ (in units $2\pi/L$): $k_x=h-\bar{h}$ or $k_x=h-\bar{h}+L/2$.
 The system size $L$ corresponds to the number of plaquettes of the basis, also denoted by $L$ in the previous and upcoming
 discussion.
 By rescaling the eigenenergies
 obtained from exact diagonalization according to Eq.~(\ref{CFT_energy_levels}),
  we are able to identify a number of conformal field theories.

\begin{figure}[t]\begin{center}
\includegraphics[width=8.8cm]{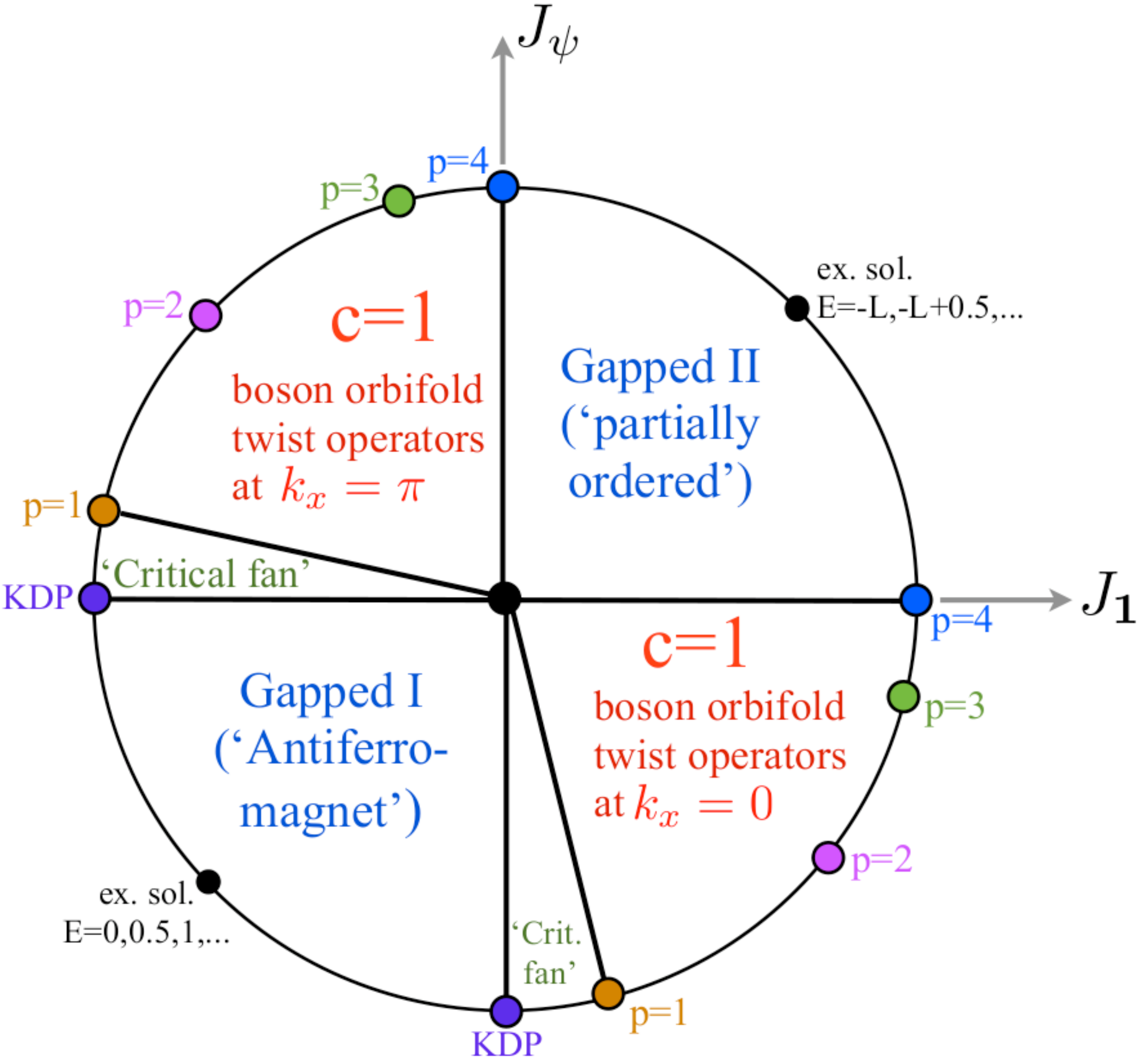}
\caption{Half-integer sector (HIS): Phase diagram at angles $\theta=\pi/4$, and $\phi\in[0,2\pi)$\cite{angles2}. The coupling constants are
$J_{\tr}=\cos(\phi)$ and  
$J_{\psi}=\sin(\phi)$. 
 The positions of some of the $c=1$ theories 
  (boson orbifold compatified on a circle of radius  $R=\sqrt{2p}$) are indicated, see 
	  table~\ref{comp_boson_table} for details on the operator content.
   For example, $p=4$ stands for the $4$-state Potts CFT. 
  }
\label{PD_theta25_phi_HIS}
\end{center}\end{figure}
		

\subsection{$\mathbb{Z}_2$ orbifold of the boson compactified on a circle of radius $R=\sqrt{2p}$}
\label{orbifold}
As was mentioned in the introduction, we identify a number of conformal
 field theories with central charges $c=1$. We observe that these theories are part of the $\mathbb{Z}_2$ orbifold the the
  bosonic theory compactified on a circle of radius $R=\sqrt{2p}$, with each integer parameter
   $p\ge 1$ defining a rational CFT \cite{ginsparg,dijkgraaf}. 
Aside from the ground state ($h=\bar{h}=0$),
 there are two  fields with scaling dimension $h_0+\bar{h}_0=1/8$, two fields with scaling dimension $9/8$ 
(these four operators are the so-called twist operators),  
one field with scaling dimension $2$, two fields with scaling dimension $p/2$, 
   and $p-1$ fields with scaling dimensions $n^2/2p$, $n=1,2,...,p-1$ (see Table~\ref{comp_boson_table}).
   The more prominent of the critical theories of the $\mathbb{Z}_2$ boson orbifold are 
    the Kosterlitz-Thouless theory ($p=1$), the theory of two decoupled Ising models ($p=2$), the
	  $c=1$ parafermion CFT ($p=3$), the $4$-state Potts model ($p=4$), and the $c=1$ superconformal CFT ($p=6$).
  A number of  the orbifold theories (those with $p\le 4$) are  observed  in a critical line of
   the Ashkin-Teller
    model \cite{Ashkin_Teller} which is  a two-dimensional classical lattice model of
	 two decoupled Ising models which are coupled by a four-spin interaction.
	
	\begin{figure}[t]
\begin{center}\includegraphics[width=9.2cm]{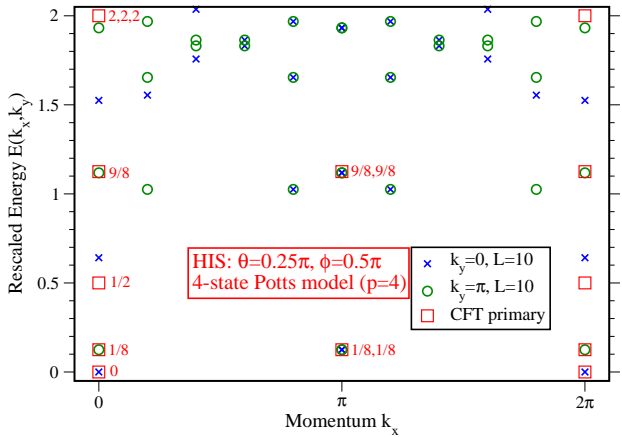}
\caption{HIS: Rescaled energy spectrum (from exact diagonalization at system size $L=10$) at  $\theta=\pi/4$, $\phi=\pi/2$,
 and the CFT assignments of the   4-state Potts model. There are three fields with scaling dimension $2$ (see Table~\ref{comp_boson_table}),
 at momentum $k_x=0$, however, the  finite-size effects are rather strong. }
 \label{HIS_4state_potts_Z2}
 \end{center}\end{figure}
	
		   	   There exists a  relation between the $c=1$ orbifold theories and  the extended Dynkin diagrams $\hat{D}_n$
    of the simply-laced affine Lie-algebras of type $D$: for $p=m^2$, $m=1,2,...$, 
	 the corresponding
	$\mathbb{Z}_2$ orbifold theory is associated with the extended Dynkin diagram $\hat{D}_{\sqrt{p}+2}$ \cite{ginsparg}.
The extended Dynkin diagrams $\hat{D}_n$ define so-called 
 restricted-solid-on-solid (RSOS) models which  are 2D statistical lattice models 
	 whose degrees of freedom  are integer-valued heights on the nodes of the lattice
	  with the constraint that heights on nearest-neighbouring
	   lattice sites are adjacent nodes in the defining Dynkin diagram.
 The partition function of these RSOS models is a discrete version of the partition function of the rCFTs associated 
  with the respective Dynkin diagram   \cite{Pasquier,Kuniba}.


\subsection{Numerical results in the half-integer sector (HIS): $|J_r|=|J_p|$  }\label{Num_HIS}
In this section we discuss the results of the  exact diagonalization of the Hamiltonian (\ref{Ham_HIS}) 
 for equal magnitude of plaquette and rung terms, i.e., $|J_p|=|J_r|$.
  In the following, we refer to the case $\theta=\pi/4$, however, all other cases of equal strength of rung and plaquette coupling 
  ($\theta=3\pi/4$, $\theta=5\pi/4$, $\theta=7\pi/4$) yield the identical results\cite{angles2}.

\begin{figure}[t]
\begin{center}\includegraphics[width=9.2cm]{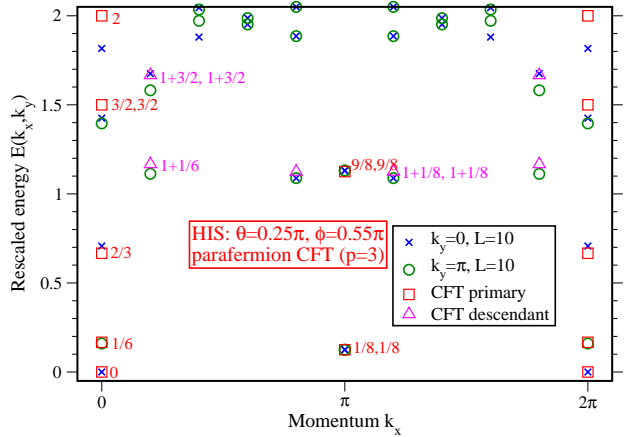}
\caption{HIS: Rescaled energy spectrum (from exact diagonalization) at  $\theta=\pi/4$, $\phi=0.55\pi$,
 and the CFT assignments of the parafermion CFT with $c=1$.
}
\label{HIS_para_Z2}
\end{center}\end{figure}

\subsubsection{Gapless phases}\label{HIS_gapless}
At angles $\phi=0$ and $\phi=\pi/2$, the model is critical and described by the $4$-state Potts model, 
 as shown in Fig.~\ref{HIS_4state_potts_Z2}, which is confirmed by an
  exact solution (see section~\ref{exact_solutions}). In fact, for all angles 
   $\phi \in [-\pi/2,0]$ and $\phi\in [\pi/2,\pi]$ the system is gapless with continuously varying critical exponents.
 We are able to match the  energy spectra at different angles $\phi$ to  
  several of the  orbifold theories, as indicated in the phase diagram Fig.~\ref{PD_theta25_phi_HIS} (see
 also a figure of the $c=1$ parafermion CFT Fig.~\ref{HIS_para_Z2}). 
 The  two gapless phases in the phase diagram (Fig.~\ref{PD_theta25_phi_HIS}) differ by the
  $k_x$-momentum quantum numbers     of the twist operators. In one of the  gapless phases, the four twist operators
   with scaling dimensions $1/8$, $1/8$, $9/8$, $9/8$ have momentum quantum numbers
    $(k_x,k_y)=(0,0),(0,\pi),(0,0),(0,\pi)$, while in the other gapless phase the momenta
	are  $(k_x,k_y)=(\pi,0),(\pi,\pi),(\pi,0),(\pi,\pi)$. 
	The eigenenergies associated with the remaining operators always appear in momentum sector $k_x=0$.
	 The   fields with scaling dimensions 
	 $n^2/2p$, $n=1,...,p-1$, have momentum quantum numbers $k_y=0$ if $n$ is even, and $k_y=\pi$ if $n$ is odd.
The numerical results indicate that 
	fields with scaling dimensions $p/2$ are both in momentum sector $k_y=0$ for $p$ even, while for 
	$p$ odd, these two fields have momentum numbers $k_y=0$ and $k_y=\pi$, respectively.
	The marginal operator is in momentum sector $(k_x,k_y)=(0,0)$.
	We located the critical theories at angles $\phi=0$ and $\phi=0.5\pi$ 
	($p=4$),
	 $\phi\approx 0.55\pi$ and $\phi\approx 1.95\pi$   ($p=3$), 
	  $\phi\approx 0.75\pi$ and $\phi\approx 1.75\pi$ ($p=2$), and
 $\phi\approx 0.95\pi$ and $\phi\approx 1.55\pi$ ($p=1$).
The  exact locations of these orbifold theories can be determined by comparison with\cite{Kohmoto}.
 The Hamiltonian (\ref{Ham_HIS}) is integrable along the orbifold line, i.e., between the
  Kosterlitz-Thouless point and the $4$-state Potts point (see \cite{Kohmoto} and references therein).

  The  region between the Kosterlitz-Thouless transition ($p=1$) and the points $\phi=\pi$ and $\phi=3\pi/2$, respectively,
  is gapless with continuously varying critical exponents. This  region is  denoted as `critical fan' in\cite{Kohmoto}.
   The transition points $\phi=\pi$ and $\phi=3\pi/2$ are gapless, and exhibit a 
   three-fold degenerate ground state [momenta $(k_x,k_y)=(0,0),(\pi,0),(\pi,\pi)$ at $\phi=\pi$, and
    $(k_x,k_y)=(0,0),(0,\pi),(\pi,\pi)$ at $\phi=3\pi/2$]. They mark a first order transition between the critical phases
	 and gapped phase I, as indicated by a  
	jump of the ground state energies (as a function of $\phi$) at these angles (not shown here).
These two critical points are in the universality class of the potassium
	 dihydrogen phosphate (KDP) model\cite{Lieb}.

\subsubsection{Gapped phases}\label{gapped}
We briefly discuss the two gapped phases in the phase diagram Fig.~\ref{PD_theta25_phi_HIS}.
In gapped phase I [$\phi\in(\pi,3\pi/2)$], the ground state
    is two-fold degenerate. Above the ground state, a flat  quasiparticle band is observed.
At angle $\phi=5\pi/4$, the Hamiltonian is of
	  form $H=-\frac{1}{2}\sum_{i} P_{2i-1}^{(\sigma)}-\frac{1}{2}\sum_i R_{2i}^{(\sigma)}$. At this point, 
	  the energy is minimized if all rungs  have occupation $\sigma$.
	  This is realized for any configuration of form [we omit the indices $c_i=\sigma$, i.e., $|\Psi\rangle 
	  = |(a_1,b_1),(a_3,b_3),...\rangle$]
	$   |\Psi_{I}\rangle= |(a_1, \sigma),(\sigma,b_3),(a_5,\sigma),...\rangle $, where $a_1,b_3,a_5,...\in \{\tr,\psi\}$.
All states of this form, and hence also the ground states,
 appear only in momentum sectors $(k_x,k_y)=(0,0)$ and $(k_x,k_y)=(\pi,\pi)$.
The ground states are the product states 
of local states of form
$|(a_i,b_i)\rangle = \frac{1}{\sqrt{2}}(|(\tr,\sigma)\rangle 
-|(\psi,\sigma)\rangle)$ ($i=1,5,...2L-3$), 
 and $|(a_{i},b_{i})\rangle = \frac{1}{\sqrt{2}}(|(\sigma,\tr)\rangle 
-|(\sigma,\psi)\rangle)$ ($i=3,7,...,2L-1$),
 and they are hence a superposition of all states of form $|\Psi_{I}\rangle$, where the magnitude of the  weights depends on the 
  multiplicities of the states according to the symmetries.
   The numerical results confirm that this is indeed the correct construction for any 
	  point in the gapped phase I.
The  symmetry and the exact form of the ground states is indicative of a system of two independent sublattices.
 Gapped phase I corresponds to the `antiferromagnetic frozen phase' in \cite{Kohmoto}.

In gapped phase II, the ground state
 is also two-fold degenerate, and the 
 quasiparticle dispersion has a leading {\it cosine} shape.
At coupling parameter 
 $\phi=\pi/4$, the Hamiltonian is of
	  form $H=\frac{1}{2}\sum_{i} P_{2i-1}^{(\sigma)}+\frac{1}{2}\sum_i R_{2i}^{(\sigma)}$. 
	  At this point, 
	  the energy is minimized if all rungs (in the ladder basis) have occupation  $\tr$ or $\psi$.
	  This is realized for any configuration of form 
	 $ |\Psi_{II}\rangle = 
 |(a_1, \sigma),(a_3,\sigma),(a_5,\sigma),...\rangle$ where 
  $a_1,a_3,a_5,...\in \{\tr,\psi\}$.
All states of this form, and hence also the ground states,
 appear only in momentum sectors $(k_x,k_y)=(0,0)$ and $(k_x,k_y)=(0,\pi)$.
The ground states  are the product states of
 local states of form $|(a_i,b_i)\rangle = \frac{1}{\sqrt{2}} (|(\tr,\sigma)\rangle 
+|(\psi,\sigma)\rangle$), $i=1,3,..., 2L-1$
 and thus are a superposition of all states of form $|\Psi_{II}\rangle$ where the magnitude of the  weights depends on the 
  multiplicities of the states according to the symmetries. 
  This gapped phase 
 corresponds to a partially ordered phase (ordered in one of the two Ising spins)
 in \cite{Kohmoto}.

\begin{figure}[t]\begin{center}
\includegraphics[width=8.7cm]{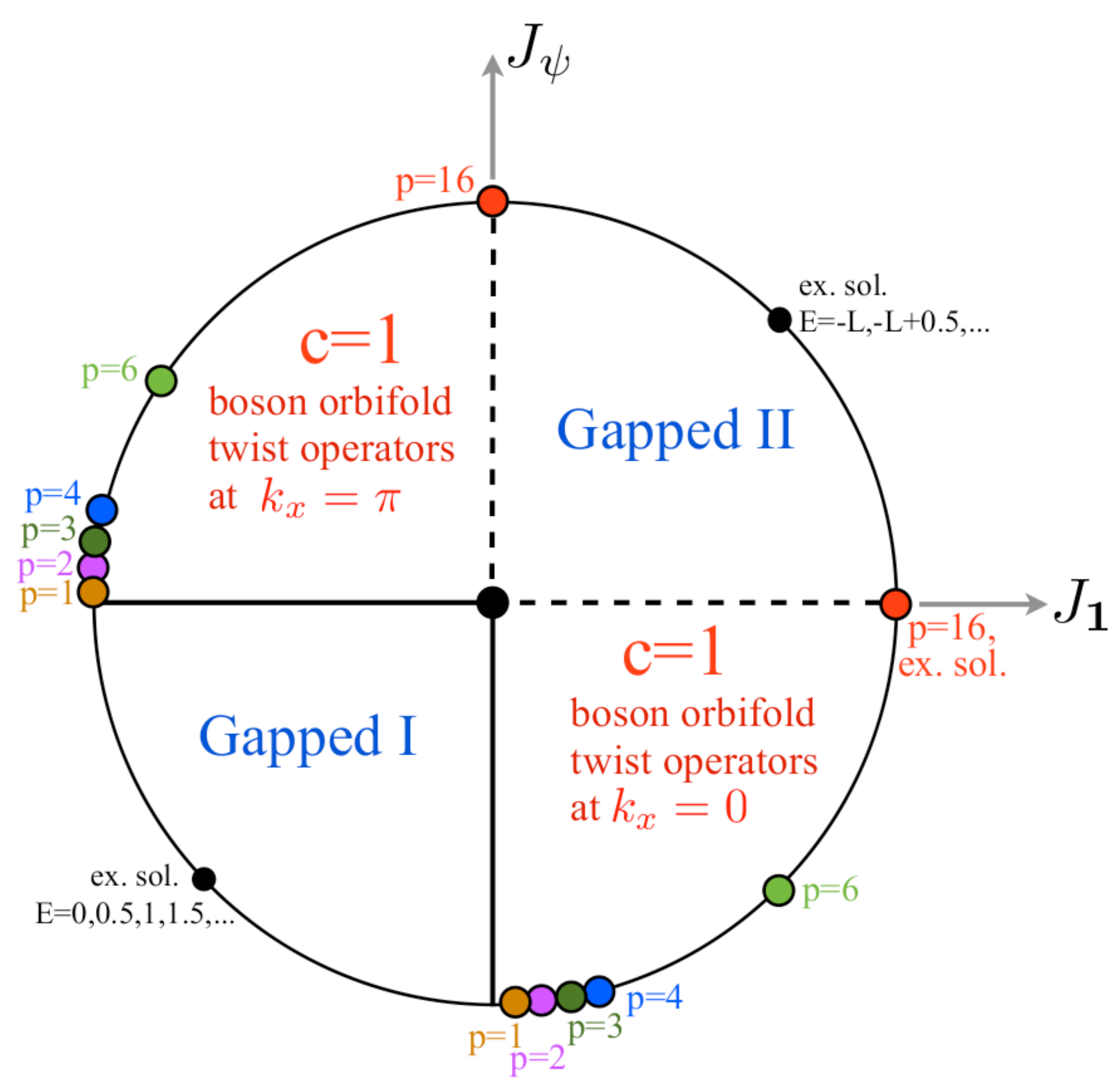}
\caption{Integer sector (IS): Phase diagram at angles $\theta=\pi/4$, and $\phi\in[0,2\pi)$\cite{angles2}. 
 The coupling constants are
$J_{\tr}=\cos(\phi)$ and  
$J_{\psi}=\sin(\phi)$. 
 The positions of some of the $c=1$ theories 
  (boson orbifold compatified on a circle of radius  $R=\sqrt{2p}$) are indicated, see 
	  table~\ref{comp_boson_table} for details on the operator content.
	  There exist exact solutions (ex. sol.) at several points: $\phi=0$, $\phi=\pi/2$ ($p=16$ boson orbifold, 
     section~\ref{Exact_IS}),  $\phi=\pi/4$, $\phi=5\pi/4$ (section~\ref{IS_gapped}).
  }
\label{PD_theta25_phi_IS}
\end{center}\end{figure}


\subsection{Numerical results in the integer sector (IS): $|J_r|=|J_p|$}
We discuss the results of the exact diagonalization of the Hamiltonian (\ref{Ham_IS})  for equal magnitude 
of plaquette and rung term. In the following we refer to the case $\theta=\pi/4$\cite{angles2}.

\subsubsection{Gapless phases}
\begin{figure}[t]
\begin{center}\includegraphics[width=9.2cm]{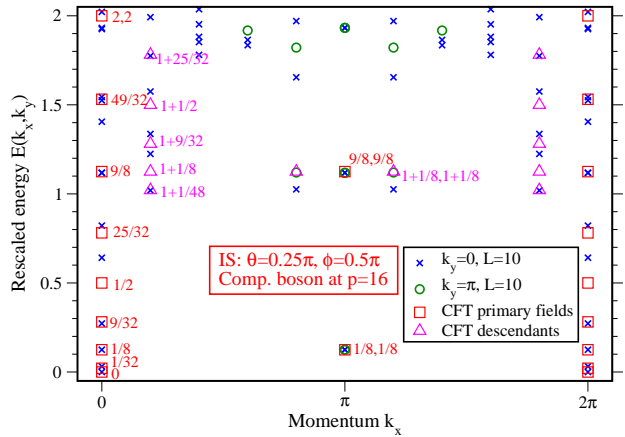}
\caption{IS: Rescaled energy spectrum (from exact diagonalization) at  $\theta=\pi/4$, $\phi=\pi/2$,
 and the CFT assignments of the $\mathbb{Z}_2$ orbifold of the compactified bosonic theory
  at radius $R=\sqrt{2p}$ with $p=16$. }
 \label{spec_IS_p16_Z2}\end{center}\end{figure}

\begin{figure}[t]
\begin{center}\includegraphics[width=9.2cm]{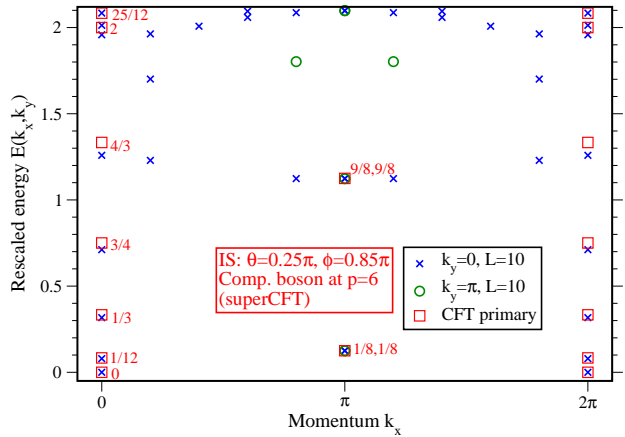}
\caption{IS: Rescaled energy spectrum (from exact diagonalization) at  $\theta=\pi/4$, $\phi=0.85\pi$,
 and the CFT assignments of the $\mathbb{Z}_2$ orbifold of the compactified bosonic theory
  at radius $R=\sqrt{2p}$ with $p=6$ ($c=1$ superconformal CFT). }
 \label{spec_IS_p6_Z2}\end{center}\end{figure}

  The phase diagram in the integer sector (IS) at equal strength of rung and plaquette term
  is similar to the one of the half-integer sector, with
   two extended gapped and two extended gapless phases, as illustrated in Fig.~\ref{PD_theta25_phi_IS}.
 However,  a much larger number of the rational $c=1$ critical theories appears.
 At angles $\phi=0$ and $\phi=\pi/2$ (where we found the $p=4$ theory 
 in the half-integer sector), we identify the $p=16$ orbifold theory, as can be seen  in Fig.~\ref{spec_IS_p16_Z2}. 
    This numerical result is confirmed by the exact solution given in section~\ref{Exact_IS}.
	We identify  the $c=1$ orbifold theories for $p=1,2,...,16$ in the two gapless phases.
	 One example of the observed theories is the $c=1$ superconformal CFT ($p=6$)  which is shown in Fig.~\ref{spec_IS_p6_Z2}.
 As was the case for the half-integer sector, 
 the  two gapless phases differ by the  
  $k_x$ momentum quantum numbers  of the twist operators.
   The momentum assignments of the twist operators in the integer sector are identical to the ones in the half-integer sector.
 In contrast to the half-integer sector, all remaining fields have momentum quantum numbers $(k_x,k_y)=(0,0)$.

The angles at which some of the critical theories are located are the following: $\phi=0$ and $\phi=0.5\pi$ ($p=16$)
	 $p=16$, 
	 $\phi\approx 0.85\pi$ and $\phi\approx 1.65\pi$  ($p=6$), 
	  $\phi\approx 0.95\pi$ and $\phi\approx 1.55\pi$ ($p=4$),
	  $\phi\approx 0.97\pi$ and $\phi\approx1.53\pi$ ($p=3$),
	 	  $\phi\approx 0.985\pi$ and $\phi\approx 1.515\pi$ ($p=2$), and
 $\phi\approx 0.995\pi$ and $\phi \approx 1.505\pi$ ($p=1$).

It is difficult to determine the exact position of the transition between either of the gapless phases and 
the gapped phase I (dashed lines in Fig.~\ref{PD_theta25_phi_IS}).
However, the energy eigenvalue associated with the field with scaling dimension $9/2p$ does not become
 degenerate with the eigenvalue associated with the twist fields of scaling dimension $1/8$ 
  when approaching the gapped phase I (from either side). This means that the orbifold theory with $p=36$ does not 
  appear, and 
 thus the orbifold theory with  $p$  must be one of the theories  with $16\le p<36$.
From the analytical results applicable to the half-integer sector (section~\ref{AT_HIS}) it is known that the
 transition between gapped phase I and the critical phases is located at $\phi=0$ and $\phi=\pi/2$, respectively.
 It therefore seems likely that the 
  corresponding  transition in the integer sector is also located at these angles.

\subsubsection{Gapped phases}\label{IS_gapped}
We very briefly remark on the gapped phases of phase diagram Fig.~\ref{PD_theta25_phi_IS}.
In gapped phase I, a  flat quasiparticle band is observed above a 
highly degenerate ground state. These degenerate ground states are 
superpositions of states where local variables $(a_i,b_i)=(\sigma,\sigma)$ 
 are followed by variables $(a_{i+2},b_{i+2})=(\tr,\tr)$, $(a_{i+2},b_{i+2})=(\psi,\psi)$, 
 $(a_{i+2},b_{i+2})=(\psi,\tr)$ or $(a_{i+2},b_{i+2})=(\tr,\psi)$, 
  and vice versa.
In gapped phase II, the ground state is non-degenerate, and the quasiparticle band exhibits a leading {\it cosine} shape.

\subsection{Phase diagram for $|J_r|\ne |J_p|$}  \label{away_from_rp}
The equivalence of the half-integer sector model (\ref{Ham_HIS})
   with the Ashkin-Teller
   model yields the  phase diagram of the half-integer sector (for details of the 
    various phases in spin language see \cite{Kohmoto}).
	      We numerically verified the existence of a critical point which is described by the  Ising CFT with central charge $c=1/2$ 
	 for coupling parameters $\theta=0.32\pi$ and $\phi= 0.352\pi$, which is in agreement with
	  prior results on the quantum Ashkin-Teller model \cite{Kohmoto,Gehlen}. Each  of the three primary fields of the Ising theory 
	  appears twice,   where momentum symmetry sectors are $(k_x,k_y)=(0,0),(0,\pi)$ for operators with scaling dimensions $0$ and $1$, 
	   and $(k_x,k_y)=(\pi,0),(\pi,\pi)$ for operators with scaling dimension $1/8$.

In the integer sector, the exact diagonalization of Hamiltonian (\ref{Ham_IS}) at coupling parameters $\theta=0.32\pi$, $\phi= 0.352\pi$
 yields that the corresponding phase is  of Ising universality, too.
However,  the field with scaling dimension $1/8$ appears with triple degeneracy [momenta $(k_x,k_y=(0,0),(\pi,0),(\pi,\pi)$], while the fields
 with scaling dimensions $0$ and $1$ appear only once [both at momentum $(k_x,k_y)=(0,0)$].
Further details of the phase diagram in the integer sector away from $|J_r|=|J_p|$ remain to be studied.
	    
\section{Exact solutions at critical points: Dynkin diagrams}\label{exact_solutions}
 In this section, we 
  identify  the Hilbert space of our model with certain
   extended Dynkin diagrams. We observe that the Hamiltonian, 
   at certain coupling constants, corresponds to the restricted-solid-on-solid models which are associated with these
  extended Dynkin diagrams.

 \begin{figure}[t]
\begin{center}
\includegraphics[width=3.2cm]{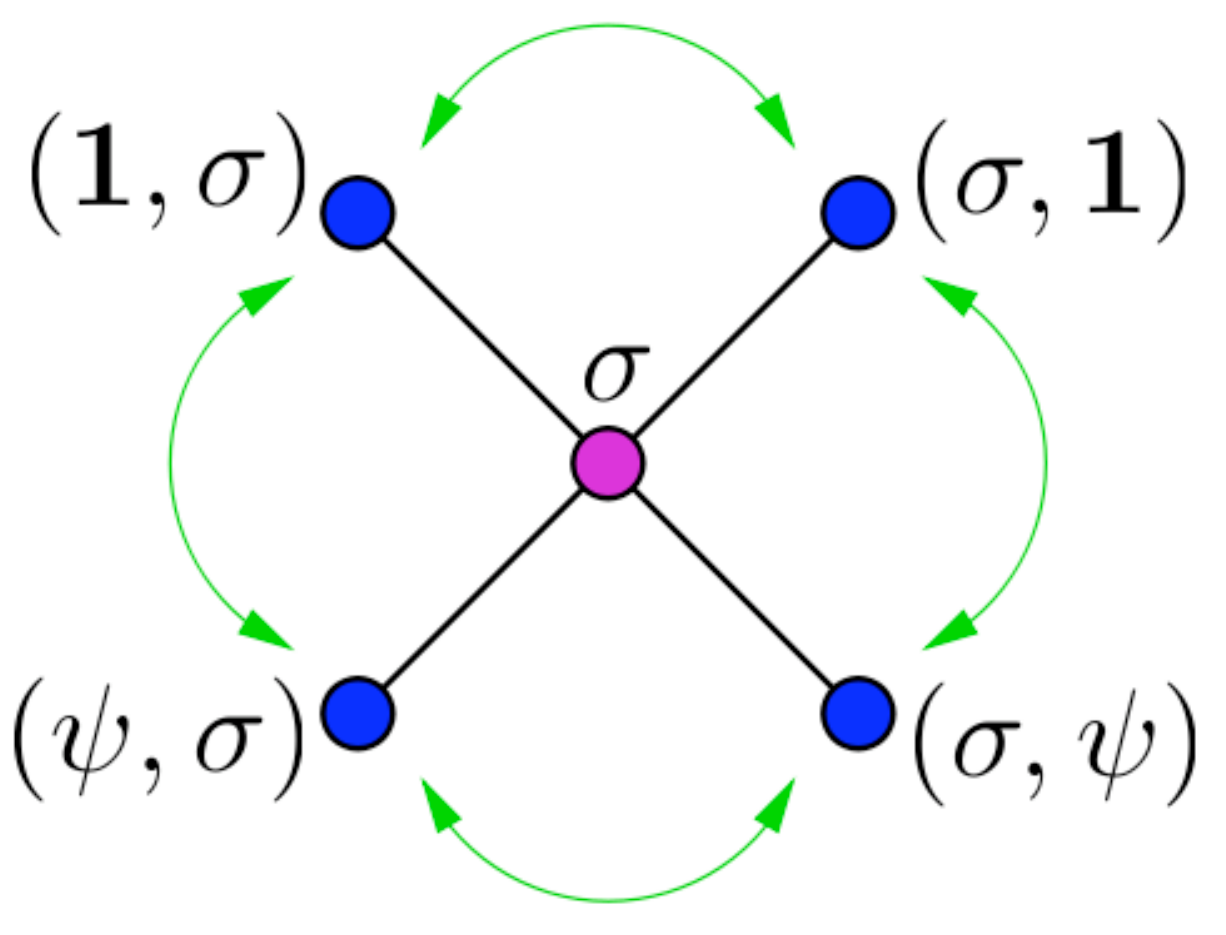}
\caption{$\hat{D}_4$ Dynkin diagram associated with the Hilbert space of the half integer sector (HIS).
The symmetries of the Hamiltonian (\ref{Ham_HIS}) under exchange of variables are indicated by green arrows. }\label{dynkin_d4}
\end{center}
\end{figure}

\subsection{Half-integer sector (HIS)}\label{Exact_HIS}

We associate a  label $c_i=\sigma$ (terminology of Fig.~\ref{basis_labeling}) with the even numbered `sites' $i$.
With the odd-numbered `sites' $i$
we associate
a variable consisting of a pair of labels, $(a_i,b_i)$, which 
can assume four values, i.e., $(a_i,b_i)=(\tr,\sigma)$, $(a_i,b_i)=(\sigma,\tr)$, $(a_i,b_i)=(\psi,\sigma)$, and $(a_i,b_i)=(\sigma,\psi)$.
If variables $(a_i,b_i)$ and $c_{i\pm 1}$ are allowed to meet at the vertices (as a consequence of the fusion rules)
 they are adjacent nodes on the Dynkin diagram of the extended $\hat{D}_4$ Lie algebra,
as illustrated in Fig.~\ref{dynkin_d4}.
Any  local label $(a_i,b_i)$
at an odd-numbered site $i$  allows for label $c_{i-1}=\sigma$
at the neighboring even-numbered sites,
which is reflected in the fact that label $\sigma$ is connected by a line all four possible labels
 $(a_i,b_i)$  in the Dynkin diagram Fig.~\ref{dynkin_d4}.

 The adjancency matrix\cite{adjacency_matrix} of  the $\hat{D}_4$
 Dynkin diagram of Fig.~\ref{dynkin_d4} is given by 
\begin{equation}
A_{\hat{D}_4} = \left ( \begin{array}{ccccc} 0&0&1&0&0\\0&0&1&0&0\\1&1&0&1&1\\0&0&1&0&0\\0&0&1&0&0\end{array}\right )\, ,
\end{equation}
where the matrix indices are associated with the five different variables
 in the following order: $(\sigma,\psi),(\psi,\sigma),\sigma,(\sigma,\tr),(\tr,\sigma)$.
The largest eigenvalue of $A_{\hat{D}_4}$ is $2$, and the corresponding eigenvector is given by 
\begin{equation}
v = (v_{(\sigma,\psi)},v_{(\psi,\sigma)},v_{\sigma}, v_{(\sigma,\tr)},v_{(\tr,\sigma)}) = (1,1,2,1,1)\, .
\end{equation}
 The operators
 \begin{eqnarray}
&&  e_i|x_1,...,x_{i-1},x_i,x_{i+1},...x_{2L}\rangle \label{dynkin_rep}  \\
 &=& \sum_{x_i'}  [(e_i)_{x_{i-1}}^{x_{i+1}}]_{x_i}^{x_i'} |x_1,...,x_{i-1},x_i',x_{i+1},...,x_{2L}\rangle, \nonumber \\[1mm] 
&&  \;\;  [(e_i)_{x_{i-1}}^{x_{i+1}}]_{x_i}^{x_i'}= \delta_{x_{i-1},x_{i+1}} \sqrt{
  \frac{v_{x_i}v_{x_i'}}{v_{x_{i-1}}v_{x_{i+1}}}}\, ,
  \nonumber 
 \end{eqnarray} 
form a representation of the 
Temperley-Lieb algebra \cite{Temperley_Lieb},
\begin{eqnarray}
e_i^2 &=& \mathcal{D} e_i\, ,\\ \nonumber
e_ie_{i\pm 1}e_i &=& e_i\, , \\ \nonumber
[e_i,e_j]&=&0\;\;{\rm for}\;\; |i-j|\ge 2\, ,
\label{TL_alg}
\end{eqnarray}
 where $\mathcal{D}=2$.
 At coupling constants $\theta=\pi/4$, $\phi=0$\cite{angles2}, the local terms of our Hamiltonian Eq.~(\ref{Ham})
 (in the half-integer sector) $H_i=P_i^{(\tr)}$
  ($i$ odd) and $H_i=R_i^{(\tr)}$ ($i$ even) equal to 
 $H_i=-\frac{1}{2}e_i$  which can be seen by evaluating the operators $e_i$.
 It can be shown that the Hamiltonian (at these coupling constants) defines the two-row transfer matrix 
  of the RSOS model that is associated with the $\hat{D}_4$ Dynkin diagram \cite{transfer_matrix}.
   Consequently, our model in the HI sector at angles $\theta=\pi/4$, $\phi=0$\cite{angles2} 
   is described by the $4$-state Potts CFT \cite{Pasquier,ginsparg}.

\begin{figure}[t]
\begin{center}
\includegraphics[width=5.5cm]{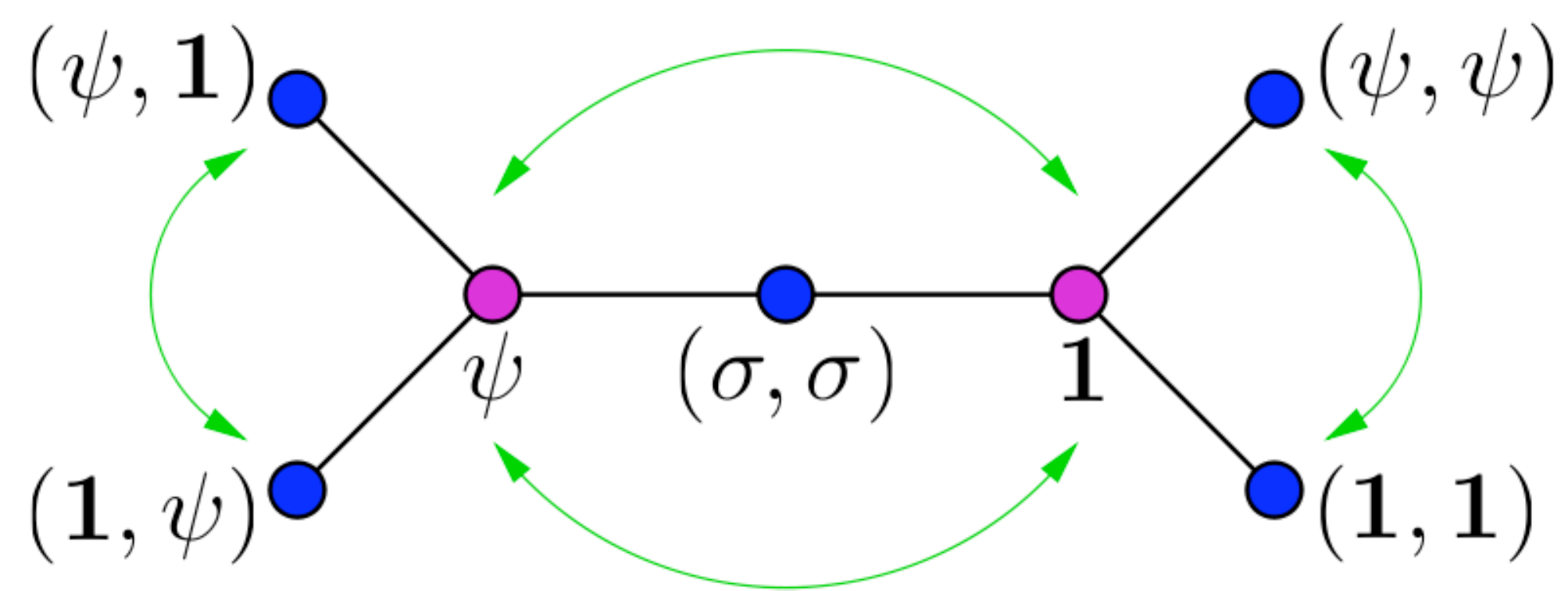}
\caption{$\hat{D}_6$ Dynkin diagram associated with the Hilbert space of the integer sector (IS).
The symmetries of the Hamiltonian (\ref{Ham_IS}) under exchange of labels are indicated by green arrows.}\label{dynkin_d6}
\end{center}
\end{figure}

\subsection{Integer sector (IS)}\label{Exact_IS}

In analogy to the discussion of section~\ref{Exact_HIS}, 
we associate a  label $c_i=\tr$ or $c_i=\psi$ with the even numbered sites, while the  odd-numbered sites 
are associated
with variable consisting of a pair of labels, $(a_i,b_i)$ which 
can assume five values, i.e., $(a_i,b_i)=(\tr,\tr)$, $(a_i,b_i)=(\sigma,\sigma)$, $(a_i,b_i)=(\psi,\psi)$, $(a_i,b_i)=(\psi,\tr)$
and $(a_i,b_i)=(\tr,\psi)$.
Variables $(a_i,b_i)$ and $c_{i\pm 1}$  that may fuse  at the vertices  [according to the fusion rules Eq.~(\ref{fusion_rules})]
are adjacent nodes on the Dynkin diagram of the extended $\hat{D}_6$ Lie algebra,
as illustrated in Fig.~\ref{dynkin_d6}.
For example, a local label $(a_i,b_i)=(\sigma,\sigma)$
at an odd-numbered site $i$  allows for labels $c_{i\pm 1}=\tr$ and $c_{i \pm 1}=\psi$
at the neighboring even-numbered sites,
which is reflected in the fact that label $(\sigma,\sigma)$ is connected by a line to both labels
$\tr$ and $\psi$ in the Dynkin diagram.
The components $v_{x_i}$ of the eigenvector associated with the largest eigenvalue of the adjancency matrix of the 
$\hat{D}_6$ diagram 
  define a representation Eq.~(\ref{dynkin_rep})  of the Temperley-Lieb algebra 
 associated with the $\hat{D}_6$ diagram.
  Again, it is straightforward to verify that the Hamiltonian  in the 
   integer sector at parameters $\theta=\pi/4$, $\phi=0$\cite{angles2} 
 is of form $H=-\frac{1}{2}\sum_i e_i$. 
 This means that the Hamiltonian (at these coupling parameters) is that of the RSOS model defined by
  the $\hat{D}_6$ Dynkin diagram, and hence the critical theory is the $p=16$ 
 $\mathbb{Z}_2$ boson orbifold theory  \cite{ginsparg}.


\section{Conclusions}
 We study a quantum double model whose degrees of freedom are Ising anyons, and whose Hamiltonian 
implements a competition between single and double topologies.
 We observe a series of quantum critical points described by conformal field theories with central charge $c=1$.
 These critical theories are part of the $\mathbb{Z}_2$ orbifold of the bosonic theory compactified on a circle.
  By associating the Hilbert space of our model with certain extended Dynkin diagrams,
   we find exact solutions of our model at some critical points. In one of its Hilbert space sectors, our model 
    corresponds to the quantum Ashkin-Teller model.
 
  This work demonstrates the exciting physics of quantum double models which are of great interest in the
  context of topologically ordered phases of matter and topological quantum computation.
   It contributes further to the understanding of models of
  interacting non-abelian anyons \cite{Feiguin_07,Trebst_07,anyon_chains,goldenladder,anyon_review}.
  
      The author thanks the anonymous referee for useful comments and  E. Ardonne,
	   D. Huse, A. Kitaev, A. Ludwig, S. Trebst, M. Troyer and Z. Wang 
 for   enjoyable collaboration on related work.



\begin{thebibliography}{10}
\bibitem{Fibonacci}
The Fibonacci theory is a  simple class of non-abelian anyons, which  contains only one
 anyon species, the Fibonacci anyon $\tau$. For more details, see, for example,  \cite{review_anyons,anyon_review}.
\bibitem{angles2}
The choice $\theta=\pi/4$ and $\phi\in [0,2\pi)$ is equivalent to 
 the choices $\theta=3\pi/4$ and $\phi\leftarrow \pi - \phi$, $\theta=5\pi/4$ and 
 $\phi \leftarrow \pi+\phi$, as well as $\theta=7\pi/4$ and $\phi \leftarrow 2\pi -\phi$.
\bibitem{adjacency_matrix}
The adjacency matrix $A$ is the matrix whose only non-zero  matrix elements  are 
$A_{\alpha,\alpha'}=1$ when $\alpha$ and $\alpha'$ are nearest neighbors
on the Dynkin diagram.
\bibitem{transfer_matrix}
We do not present this calculation here.
However, it proceeds analogously to the one given in \cite{Feiguin_07}.
\end{thebibliography}
\end{document}